\documentclass[letterpaper, 10 pt, onecolumn, conference]{ieeeconf}
\IEEEoverridecommandlockouts                              

\usepackage{hyperref}
\usepackage{amssymb}
\usepackage{mathtools}
\usepackage{multirow}
\usepackage{subcaption}
\usepackage{booktabs}
\usepackage{float}
\usepackage{tikz}
\usepackage{collcell}
\usepackage{color, colortbl}
\definecolor{Gray}{gray}{0.9}
\definecolor{high}{HTML}{32D732} 
\definecolor{low}{HTML}{F6B914} 
\newcommand*{\opacity}{40}
\newcommand*{\minval}{0.42}
\newcommand*{\maxval}{0.98}
\newcommand\vertarrowbox[3][6ex]{%
  \begin{array}[t]{@{}c@{}} #2 \\
  \left\uparrow\vcenter{\hrule height #1}\right.\kern-\nulldelimiterspace\\
  \makebox[0pt]{\scriptsize#3}
  \end{array}%
}
\newcommand{\gradient}[1]{
    \ifdimcomp{#1pt}{>}{\maxval pt}{#1}{
    \ifdimcomp{#1pt}{<}{\minval pt}{#1}{
         \pgfmathparse{int(round(100*(#1/(\maxval-\minval))-(\minval*(100/(\maxval-\minval)))))}
        \xdef\tempa{\pgfmathresult}
        \cellcolor{high!\tempa!low!\opacity} #1
    }}
 }

\title{\LARGE \bf Metrics for 3D Object Pointing and Manipulation in Virtual Reality}

\author{Eleftherios Triantafyllidis, Wenbin Hu, Christopher McGreavy and Zhibin Li
\thanks{The authors are with the School of Informatics, The University of Edinburgh, Edinburgh, United Kingdom.}%
\thanks{{\tt\small \{eleftherios.triantafyllidis,wenbin.hu,c.mcgreavy,zhibin.li\}@ed.ac.uk}}}

\begin{document}
\maketitle
\thispagestyle{empty}
\pagestyle{empty}

\begin{figure}[H]
\centering
\vspace{-6mm}
  \includegraphics[width=1.0\textwidth]{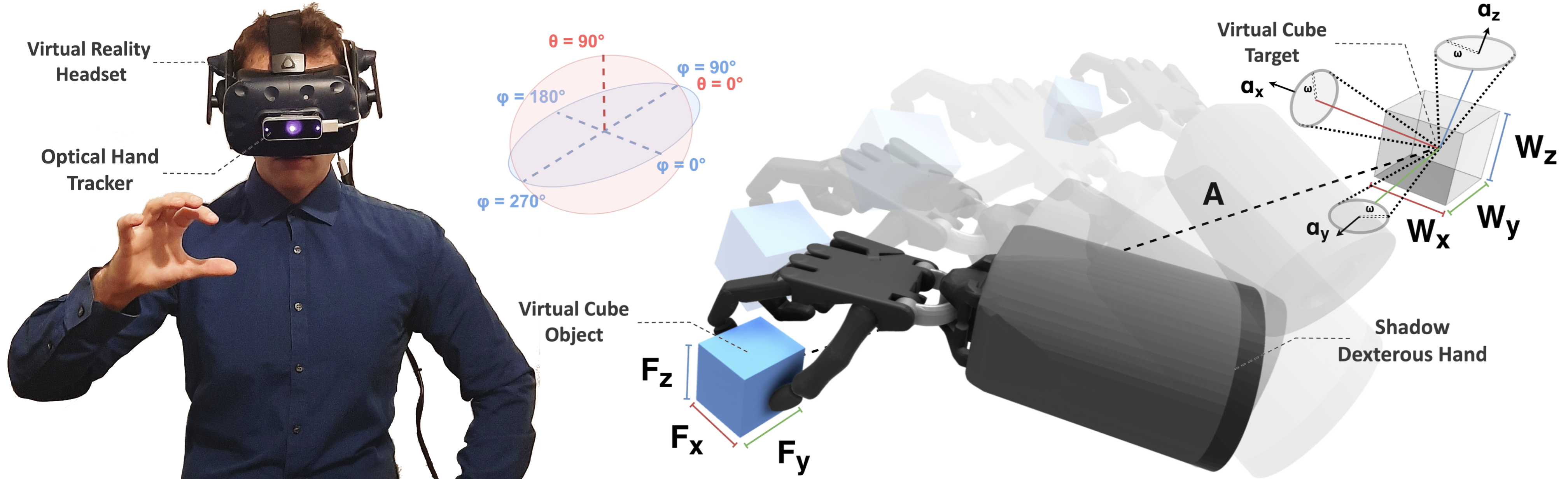}
  \caption{An operator interacts with objects in full 3D virtual reality with all task-related spatial variables.}~\label{figure:teaserImage}
  \vspace{-4mm}
\end{figure}

\begin{abstract}
Assessing the performance of human movements during teleoperation and virtual reality is a challenging problem, particularly in 3D space due to complex spatial settings. Despite the presence of a multitude of metrics, a compelling standardized 3D metric is yet missing, aggravating inter-study comparability between different studies. Hence, evaluating human performance in virtual environments is a long-standing research goal, and a performance metric that combines two or more metrics under one formulation remains largely unexplored, particularly in higher dimensions. The absence of such a metric is primarily attributed to the discrepancies between pointing and manipulation, the complex spatial variables in 3D, and the combination of translational and rotational movements altogether. In this work, four experiments were designed and conducted with progressively higher spatial complexity to study and compare existing metrics thoroughly. The research goal was to quantify the difficulty of these 3D tasks and model human performance sufficiently in full 3D peripersonal space. Consequently, a new model extension has been proposed and its applicability has been validated across all the experimental results, showing improved modelling and representation of human performance in combined movements of 3D object pointing and manipulation tasks than existing work. Lastly, the implications on 3D interaction, teleoperation and object task design in virtual reality are discussed.
\end{abstract}

\section{INTRODUCTION}
With the recent advances in networking and Mixed Reality (MR) technologies and in conjunction with the increasingly more immersive applications of telepresence and teleoperation, the need to measure and model human performance in 3D space has increased exponentially \cite{10.1145/3290605.3300437, 10.1145/3411763.3443442}. However, a compelling standardized metric for 3D object selection, such as those seen in Virtual Environments (VEs) and teleoperation, does not yet exist. The absence of a standardized metric severely limits inter-study comparability and most importantly transferability between the results of different studies, due to the multitude of different metrics researchers can use \cite{10.1145/1121241.1121249}. Consequently, progress towards the endeavour of a standardized formulation is still ``scattered" and often disregards important aspects that would solidify a concrete and established metric \cite{10.1145/1753846.1753867}.

To propose a higher dimensional metric for assessing human performance, we investigate Paul Fitts' original predictive model, short for Fitts' law \cite{fitts1954information, fitts1964information}. Proposed in 1954, the law has been extensively used in human-computer interaction (HCI) and ergonomics research and still represents the gold standard as a performance metric \cite{10.1145/3411763.3443442, 10.1207/s15327051hci0701_3}. This is attributed to the advantage of using Fitts' law to measure human performance in a time-based approach based on spatial data, effectively combining both time and spatial based metrics under one formulation. Fitts' law was originally formulated for 1D translational movements \cite{fitts1954information}, but has also been extended to 2D tasks \cite{10.1207/s15327051hci0701_3, doi:10.1080/00140139508925153} with its applicability highlighted in rotational tasks as well \cite{412031, 8998368, doi:10.1177/0018720810366560}. Recently, Fitts' formulation was also extended to some extent in 3D space, limited to translational tasks, with numerous reformulations \cite{10.1145/3290605.3300437, CHA2013350, murata2001extending}. Yet, these either disregarded important spatial aspects or limited their findings to very specific settings. This aggravates inter-study comparisons particularly due to variations in tested motor tasks. 

More specifically, previously proposed 3D metrics extending Fitts' law, have either disregarded combining translational and rotational tasks under one setting \cite{10.1145/3290605.3300437, CHA2013350, murata2001extending}, assessing directional \cite{8998368, doi:10.1177/0018720810366560} or inclination variations \cite{10.1145/3290605.3300437, 8998368, doi:10.1177/0018720810366560, murata2001extending} in one exhaustive study. All of the aforementioned appear to have significant effects on human performance \cite{10.1145/3411763.3443442, 10.1145/1753846.1753867}. More importantly, to date, most studies only focused on pointing tasks and by extent, there is a severely limited focus on manipulation tasks, e.g. with the incorporation of grasping and physical properties such as gravity and friction forces.

Consequently, we studied the aforementioned factors and spatial complexities seen in full 3D space to investigate their contribution towards a human performance model. More specifically, we performed an exhaustive user study in Peripersonal Space (PPS) i.e. close to the user's body, in a virtual simulation environment using a VR headset and optical hand tracking. The virtual environment consisted of simulated teleoperation tasks with high-fidelity physics leveraging the anthropomorphic dexterous robotic hand. By collectively combining and adding 3D spatial variables in four progressively more complex experiments, we derived our own model extension and compared the proposed metric with the state of the art while also verifying their applicability at each stage. 

From our results, to the best of our knowledge, we present a first of its kind 3D human performance metric based on Fitts' law, which extends beyond current work by modelling full 3D space better than existing formulations. More specifically, our metric is able to capture 3D human motion entailing combined translational and rotational movements, with varying degrees of directions and inclinations in both object pointing and manipulation under a single formulation. This metric can be used to assess human performance by modelling the complex motions, Degrees of Freedoms (DoFs) and dimensions associated with VEs entailing Virtual Reality (VR) \cite{10.1145/3290605.3300437, 9076603} as well as teleoperation \cite{10.1145/1121241.1121249, 8283715, doi:10.1177/154193128703100723, wen2020force}. Consequently, the effects of different user interfaces, devices and robotic systems on user performance can be modelled and assessed with such a metric with the added advantage of also combining time and spatial based metrics under one model.

The contributions of our work are summarized as follows:
\begin{itemize}
  \item We propose a new higher-dimensional metric to assess human motor performance in full 3D space;
  \item An intuitive motion re-targeting of hands in high-fidelity physics to allow generalization of our method towards virtual and realistic object interaction and manipulation;
  \item A thorough comparison of our proposed metric with others in the literature under different experimental settings and their validity towards higher dimensions;
  \item Study of quantities and variables to better explain the 3D spatial relationship of objects and recommendations for the design of pointing and manipulation tasks in VR and teleoperation.
\end{itemize}

In the remainder of the paper, we first introduce the most widely used extension methods based on Fitts' law and those to be compared with our metric. We then describe our methodology, apparatus and the design of the experiments which progressively add more complex spatial variables at each stage. Finally, we derive our proposed metric and discuss and compare our results with the extensions in the related work. A video presentation of this work is also available online\footnote{Video presentation of this work including the virtual tasks completed by the participants: \url{https://youtu.be/MKH2gGJQq2o}.}.

\section{RELATED WORK}
In this section, we introduce Fitts' original law for translational tasks and the current state of the art of its 2D and 3D extensions, as well as the importance of including rotation into a formulation. Ultimately, we investigate the incorporation of both translational and rotational movements under one combined model. In \autoref{table:ModelSummaryComparison} we summarize all model extensions.

\subsection{\textbf{Fitts' Original Formulation}}
Fitts' formulation has been extensively used in motion prediction in HCI and ergonomics research \cite{fitts1954information, fitts1964information}. The formulation predicts the \textit{Movement Time} (\(MT\)) of how long it takes for users to point to a target on a screen. It is formulated as:
\begin{equation}
\begin{gathered}
MT = a + b \cdot ID, \text{{where} } ID = log_2 \left ( \frac{2A}{W} \right ).
\end{gathered}
\label{eqn:fitts_law_original}
\end{equation}
Here, \(A\) represents the distance between the object and the target. \(W\) represents the width of the target area. The logarithmic term \(ID\), represents the \textit{Index of Difficulty} of the task, measured in bits per second. The resultant \(MT\) is measured in seconds. The constants \(a\) and \(b\) represent the y-intercept and slope respectively and are derived via regression analysis.

\subsection{\textbf{Importance and Limitations of Fitts' Law}}
While extensively used, the original formulation shown in Eq. \ref{eqn:fitts_law_original} suffers in terms of simplicity when full 3D space is considered. More specifically, the formulation is limited to four-key areas, namely, (i) lower-dimensional 1D and 2D space, (ii) lower DoFs entailing only translational tasks without the combination of rotational movements, (iii) pointing tasks without the addition of physical properties such as gravity or friction e.g. manipulation of objects and (iv) single line movements without the use of spatial arrangements e.g. directions and inclinations. During interactions either in VR or teleoperation, all four aspects are a fundamental and inseparable part of human motion in 3D \cite{10.1145/3411763.3443442, 8998368, 9076603, babarahmati2020robust}.

Nevertheless, the ability of the law to combine both time and spatial based metrics under a single formulation renders the pursuit of extending it to 3D of significant importance. As identified, part of the motivation of this paper stems from supplementing different evaluation metrics with a single metric in 3D space. The derivation of a model extension based on Fitts' law for full 3D space, is expected to increase the overall comparability between the results of different studies attempting to capture 3D human motion \cite{10.1145/3411763.3443442, 10.1145/1121241.1121249}, particularly attributed to the significant focus on the law in current work.

Hence, relying on multiple types of metrics to assess human performance in either VR or teleoperation, should ideally be avoided, as comparability between the results of such studies is rendered challenging \cite{10.1145/3411763.3443442, doi:10.1177/154193128703100723}. Earlier works on teleoperation \cite{doi:10.1177/154193128703100723} and recent ones with anthropomorphic robotic hands (19 DoFs) leveraging MR technologies \cite{8283715}, unfortunately, do not make use of metrics such as Fitts' law and instead rely on multiple time-/spatial-/behaviour-based metrics. As a result, comparability even between similar work is aggravated, since all compared studies need to have identical evaluation metrics.

\subsection{\textbf{2-Dimensional Extensions}}
While originally developed for 1D tasks, Fitts' predictive model has also been widely applied to 2D pointing tasks \cite{10.1207/s15327051hci0701_3, doi:10.1080/00140139508925153, welford1968fundamentals}. Hoffmann \cite{doi:10.1080/00140139508925153} conducted a series of discrete tapping tasks, using participants' fingers as pointing probes with the width of the finger added to the \(ID\) of Eq. \ref{eqn:fitts_law_original} as:
\begin{equation}
\label{eqn:hoffman}
    ID = log_2 \left ( \frac{2A}{W+F} \right ),
\end{equation}
where \(F\) represents the index finger pad size, which can be interpreted as the size of the object to be transported to the target \(W\) as a natural extension towards pick and place tasks \cite{10.1145/302979.302989}. Welford \cite{welford1968fundamentals} proposed another variant of Eq. \ref{eqn:fitts_law_original} as:
\begin{equation}
\label{eqn:welford}
    ID = log_2 \left ( \frac{A}{W} + 0.5 \right ),
\end{equation}
which has been demonstrated to do well in 2D task settings.

\noindent Mackenzie \cite{10.1207/s15327051hci0701_3} extended Fitts' original law, known as the Shannon equation, yielding a better fit and formulated as:
\begin{equation}
\label{eqn:shannon_formulation}
    ID = log_2 \left ( \frac{A}{W} + 1 \right ).
\end{equation}
The robustness and linear fit of the Shannon model has been demonstrated for both translational \cite{doi:10.1177/0018720810366560, murata2001extending} and rotational tasks \cite{doi:10.1177/0018720810366560}. The resulting \(IDs\) of Mackenzie's model are one bit less than with Fitts’s formulation.

Contrary to Fitts' model, the Shannon formulation is limited in terms of mathematical expressiveness. Though Mackenzie argued that the addition of the +1 term avoids negative \(IDs\) in the original Eq. \ref{eqn:fitts_law_original}, this is not entirely true. A negative \(ID\) in Fitts' model would mean that the cursor or probe is already within the target area. This limits theoretical justification for the purpose of higher model fitting \cite{10.1145/1753846.1753867}. Nevertheless, adopting Mackenzie's model facilitates the comparability of numerous extensions as this has become the norm for most of the subsequent models hereinafter presented.

\subsection{\textbf{3-Dimensional Extensions}}
While Fitts' law has been applied to some extent towards 3D pointing tasks \cite{CHA2013350, murata2001extending}, it does not accurately represent 3D movement \cite{10.1145/3411763.3443442, doi:10.1177/0018720810366560}. Murata and Iwase \cite{murata2001extending} introduced an extension of Fitts' law to 3D pointing tasks taking into account directional angles $\theta$ between an origin and a target. Their model is based on Eq. \ref{eqn:shannon_formulation} and formulated as:
\begin{equation}
\label{eqn:murata_inclination}
    ID = log_2 \left ( \frac{A}{W} + 1 \right ) + c \cdot \sin{\theta},
\end{equation}
with $\theta$ being the azimuth angle and effectively added to the \(ID\). Their work found that directional angles $\theta$ had an almost sinusoidal relationship with \(MT\) \cite{murata2001extending}.

A later study by Cha and Myung \cite{CHA2013350} extended the above work by adding inclination angles, representing higher dimensions in finger aimed pointing tasks in the spherical coordinate system. Based on Eq. \ref{eqn:hoffman}, it is formulated as:
\begin{equation}
\label{eqn:cha_myung_incline_azimuth}
    MT = a + b \cdot \theta_{1} + c \cdot \sin{\theta_{2}} + d \cdot log_2 \left ( \frac{2A}{W + F}\right ),
\end{equation}
where $\theta_{1}$ and $\theta_{2}$ represent the inclination and azimuth angles from the starting point to the target respectively. Directional angles followed again an almost sinusoidal relationship with \(MT\) as shown in Murata's and Iwase's work \cite{murata2001extending}. The constants \(a\), \(b\), \(c\) and \(d\) are empirically determined through linear regression. However, in the work of \cite{CHA2013350}, they limited their investigation to forward motions covering between $\pm30^{\circ}$ and $\pm60^{\circ}$ azimuth angles. 
Recent work, by Barrera Machuca and Stuerzlinger \cite{10.1145/3290605.3300437} accounted for the above by introducing pointing tasks with the use of 3D displays to point towards targets ranging from azimuth angles of \(-90^{\circ}\) to \(90^{\circ}\) and \(0^{\circ}\) to \(180^{\circ}\). Their work confirmed that left-to-right movements were easier than movements away from or towards the user. While this investigation covered a wider range of azimuth angles, it still disregarded inclination angles as the height of the objects was adjusted to the view height of each participant. However, human motor skills vary significantly with the direction of movements, for example, upward movements appear to be more demanding than downward ones \cite{CHA2013350}. Hence, it is important to study the effects of both directions and inclinations.

So far in our analysis, no model has investigated rotational variations or even combined rotational with translational movements all in one setting. Additionally, the identified spatial arrangements and factors in this section should ideally be included as well. Our work aims to address these by effectively combining them in one setting.

\begin{table*}\centering
\begin{scriptsize}
\begin{tabular}{lllccccc} \toprule
\multirow{2}{*}{\textbf{Performance Models}} & \multicolumn{2}{c}{\textbf{Model Formulation and Equation}} & \multicolumn{5}{c}{\textbf{Model Characteristics}} \\
      \cmidrule(lr){2-3} \cmidrule(lr){4-8}
 & MT (sec) & $ID$ (bit/sec) & Derived From & Space & Dir. & Inc. & Rot. \\ \midrule
 Fitts' \cite{fitts1954information} & $MT = a + b \cdot ID$ &  $ID = log_2 \left ({2A}/{W} \right )$ & [N/A] & 2D & No & No & Yes \\ \addlinespace[0.1cm]
 Hoffmann's \cite{doi:10.1080/00140139508925153} & $MT = a + b \cdot ID$ &  $ID = log_2 \left ( {2A}/{W+F} \right )$ & \cite{fitts1954information} & 2D & No & No & No\\  \addlinespace[0.1cm]
 Welford's \cite{welford1968fundamentals} & $MT = a + b \cdot ID$ & $ID = log_2 \left ({A}/{W} + 0.5 \right )$ & \cite{fitts1954information} & 2D & No & No & No\\ \addlinespace[0.1cm]
 Shannon's \cite{10.1207/s15327051hci0701_3} & $MT = a + b \cdot ID$ &  $ID = log_2 \left ({A}/{W} + 1 \right )$ & \cite{fitts1954information} & 2D & No & No & Yes\\ \addlinespace[0.1cm]
 Murata and Iwase's \cite{murata2001extending} & $MT = a + b \cdot ID$ &  $ID = log_2 \left ({A}/{W} + 1 \right ) + c \cdot \sin{\theta}$ & \cite{10.1207/s15327051hci0701_3} & 3D & Yes & No & No\\ \addlinespace[0.1cm]
 Cha and Myung's \cite{CHA2013350} & $ MT = a + b \cdot \theta_{1} + c \cdot \sin{\theta_{2}} + d \cdot ID$ &  $ID = log_2 \left ( {2A}/{W+F} \right )$ &  \cite{murata2001extending, doi:10.1080/00140139508925153} & 3D & Yes & Yes & No\\
\bottomrule
\end{tabular}
\end{scriptsize}
\caption{Summary of the models that are the most widely used to date in existing work based on Fitts' law and those compared with our proposed model. Each model's characteristics are summarized and whether important spatial variables are accounted for in the 3D domain. Dir.: Directions. Inc.: Inclines. Rot.: Rotation.}
\label{table:ModelSummaryComparison}
\end{table*}

\begin{figure*}
\centering
  \includegraphics[width=0.75\textwidth]{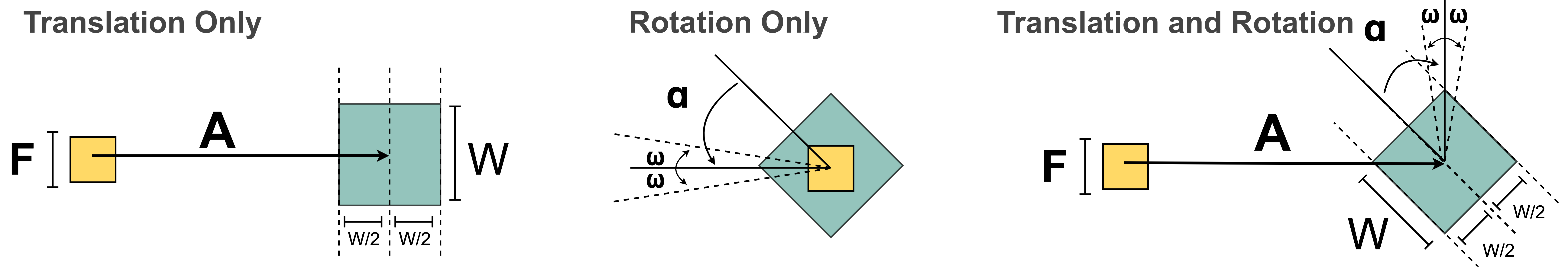}
  \caption{Illustration of three types of tasks in 2D space: translational tasks, rotational tasks and the combination of both.}~\label{2DTaskTranslationRotation}
  \vspace{-5mm}
\end{figure*}

\subsection{\textbf{Combining Translation and Rotation}}
To this point, we analyzed the most widely used extensions of Fitts' law, limited to purely translational tasks. However, while performing almost any type of manipulation or pointing task, we attempt to match the rotation of the object as well, as to satisfy certain spatial criteria \cite{10.1145/3411763.3443442, doi:10.1177/154193128703100723, Billardeaat8414}. In addition, to effectively describing general human movement, the simultaneous presence of both translation and rotation needs to be accounted for. Both are an essential and inseparable part when manipulating either real or virtual objects in the 2D and 3D domain \cite{8998368, doi:10.1177/0018720810366560, 9076603}. One early study showed that Fitts' law can be adjusted and applied to purely 2D rotational tasks, but did not investigate combined movements \cite{412031}. 

The combination of translational and rotational tasks in 2D space is visually depicted in \autoref{2DTaskTranslationRotation}. Stoelen and Akin \cite{doi:10.1177/0018720810366560} were the first to ``merge" both translational and rotational terms into an extended formulation based on Eq. \ref{eqn:shannon_formulation}. They both suggested that rotation and translation can be accounted for by effectively adding the sum of indices of difficulty for each one. The implication is that both rotation and translation are independent transformations explaining the spatial relationship of an object. The simplicity of adding both terms in their study into one index of difficulty is supported by Kulik et al. \cite{8998368}. The latter also found that long task completion times in object manipulation can be attributed to either increased cognitive workload or even to excessive accuracy requirements instructed by the researchers. Yet, the latter finding still requires further modelling of variations of accuracy requirements in order to quantify this relationship.

Nevertheless, in both cases in \cite{8998368} and \cite{doi:10.1177/0018720810366560}, combined movements and spatial arrangements were limited to 2D, i.e. following only a straight line. As future work, the authors pointed out that it is important to generalize and extend a metric to three dimensions for modelling virtual reality navigation and teleoperation \cite{doi:10.1177/0018720810366560}. Our work addresses these limitations.

\section{METHODOLOGY AND STUDY DESIGN}
In this section, we introduce our system setup including the apparatus and technologies used. We also present our hand-control approach that allowed participants to interact with the virtual objects simulated by realistic high-fidelity physics.

\subsection{\textbf{System Overview and Apparatus}}
\autoref{figure:hand_retargeting_simulation_environment} illustrates the system overview, including the simulation engine Unity3D, all necessary hardware and software. Two input sensors were used to interact with the simulation. The first sensor was the optical hand tracking device -- Leap Motion Hand Controller (LMHC). Equipped with two infrared cameras at 120Hz and with a \(135^{\circ}\) Field of View (FoV), the LMHC allowed to interface between the user's hand movements and the physics engine PhysX in Unity3D. For visual feedback, high-resolution displays were used to limit distance overestimation and degraded longitudinal control, a known issue in VEs \cite{10.1145/3290605.3300437, 10.1145/3411763.3443442}. Consequently, the Virtual Reality Head-Mounted Display (VRHMD) HTC Vive Pro was used, with a 2880 x 1600 pixel resolution display and $110^{\circ}$ FoV at 90\(Hz\). The photosensors on the VRHMD also represented our second sensor, allowing for head rotations in the VE. Furthermore, ROS\# was used to import physics models of robots and objects (Unified Robot Description Format). For all experiments, the LMHC was fixed on the front of the VRHMD.

The physics simulation time-step was set at 1000\(Hz\) to ensure robust and stable performance with realistic forces and frictions. Finally, to ensure optimal hand tracking performance, lightning conditions were consistent and operational space was limited to about 100cm as the upper maximum reaching bounds from the chest of users. In addition, a low-pass filter with a cutoff frequency of 10Hz was applied to the LMHC to reduce noise during retargeting, ensure continuity and robustness.

\subsection{\textbf{Hand Control and Input Interface}}
\begin{figure}
     \centering
     \begin{subfigure}[b]{0.51\textwidth}
         \centering
         \includegraphics[width=\textwidth]{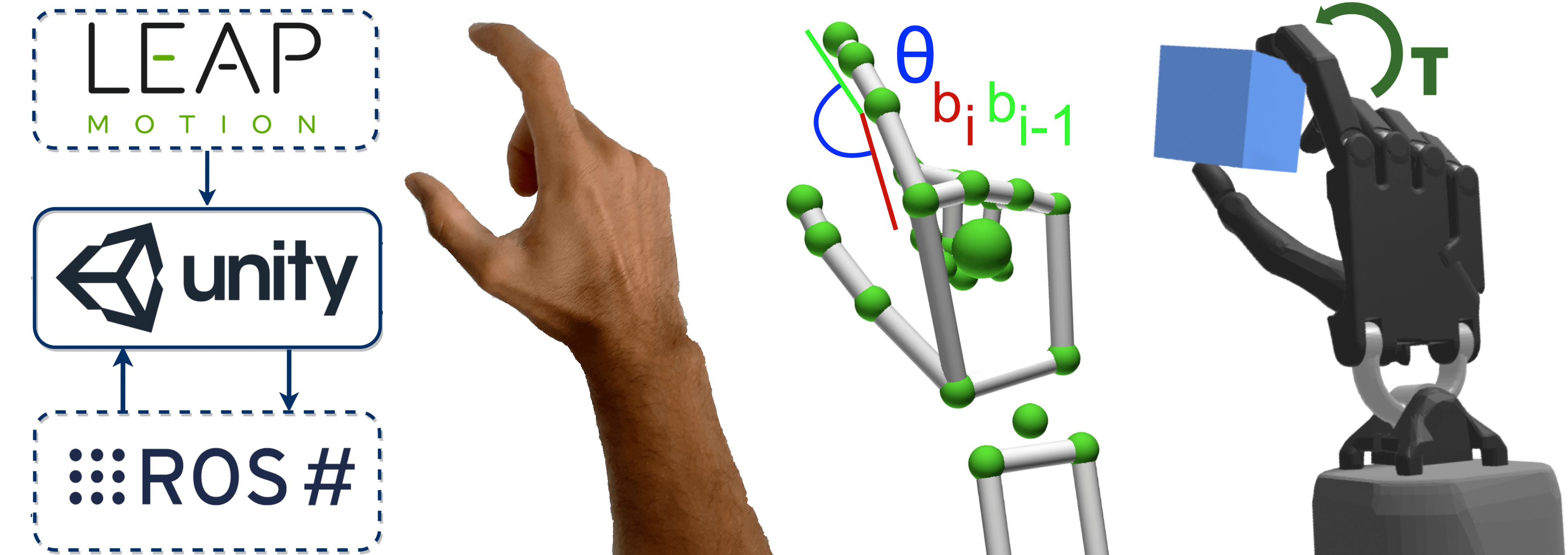}
         \vspace{-5mm}
         \caption{Physics simulation environment and hand setup.}
         \label{figure:hand_retargeting_simulation_environment}
     \end{subfigure}
     \begin{subfigure}[b]{0.49\textwidth}
         \centering
         \includegraphics[width=\textwidth]{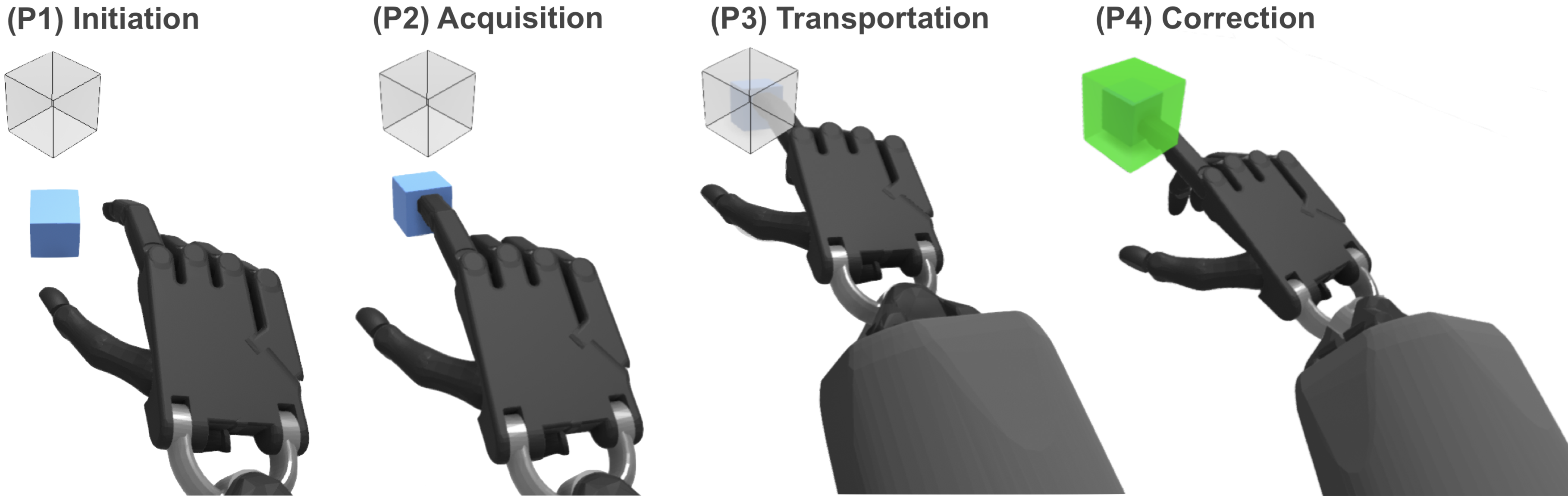}
         \vspace{-5mm}
         \caption{Pointing tasks.}
         \label{figure:PointingType}
     \end{subfigure}
     \begin{subfigure}[b]{0.49\textwidth}
         \centering
         \includegraphics[width=\textwidth]{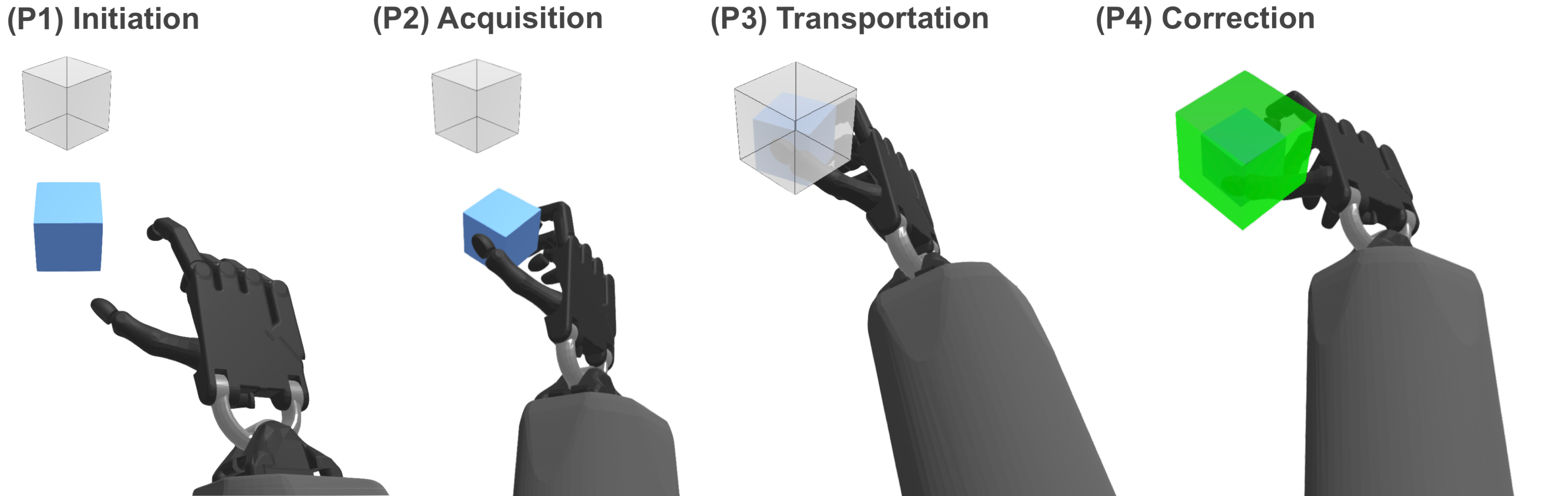}
         \vspace{-5mm}
         \caption{Manipulation tasks.}
         \label{figure:ManipulationType}
     \end{subfigure}
    \caption{The system overview and the two types of interaction in the simulation environment. (a) depicts the simulation environment and the retargeting approach: the illustration of the user's hand, the LMHC visualisation of the joint vectors and the motion retargeting of the robotic hand in Unity3D. (b) and (c) represent the pointing and manipulation tasks respectively, broken down into the four basic phases of interacting with an object. The only difference between (b) and (c) lies in phase P2, where the object is either grasped or attached to the hand depending on the type of interaction.}
\end{figure}

As shown in \autoref{figure:hand_retargeting_simulation_environment}, the user's hand movements were mapped onto an anthropomorphic Shadow Robotic hand in the simulation. The palm of the simulated hand had 6 DoF and could move freely around the virtual environment in all axes for both translation and rotation. To tele-control the virtual hand, the Cartesian hand movements and joint positions from the participant were mapped onto that in the virtual environment. The re-targeting approach was similar to the work in \cite{9076603}, where the joint positions of the user's fingers were obtained by calculating the angle $\theta$ between a joint $\overrightarrow{b_{i-1}}$ and its parent joint $\overrightarrow{b_{i}}$. The resultant angle $\theta$ from the user's finger is then incorporated in a joint PD controller to achieve the desired re-targeting joint motions from the user hand to the simulated robotic one. These are formulated as:
\begin{equation}
\begin{aligned}
\theta &= \arccos \left ( \frac{\overrightarrow{b_{i}} \cdot \overrightarrow{b_{i-1}}}{\left \| \vec{b_{i}} \right \| \left \| \overrightarrow{b_{i-1}} \right \|} \right ), \\
\tau_i &= K_{P_i} \cdot (\theta_{d_i} - \theta_{c_i}) - K_{D_i} \cdot \dot{\theta}_i,
\label{eqn:handApproach}
\end{aligned}
\end{equation}
where $\theta$ represents the desired angle for the virtual hand to match. Furthermore, $\tau$ are the torques applied to each joint \(i\). $\theta_{d_i}$ and $\theta_{c_i}$ are the desired angles from the human hand from the LMHC and the current angle of the virtual hand joint in the simulation, respectively. Finally, $\dot{\theta}_i$ is the measured velocity of the virtual hand for computing the damping torque.

Furthermore, a velocity control signal was applied to the palm of the robotic hand based on the real hand's position and orientation. Hence, the 6 DoFs of the robotic hand were controlled by matching its translation and orientation with that of the real one captured by the LMHC. The collision between the robotic hand and the objects in the simulation environment was realised via the built-in PhysX 4.0 engine in Unity3D. We furthermore retained the original joint limits as well as colliders of the virtual hand, as specified in the URDF file of the Shadow robot hand to ensure optimal and realistic actuation.

\subsection{\textbf{Task Design and Spatial Assessment}}
In each task, participants were asked to move an object from a start to a target location. Contrary to the use of a sphere or the index finger of a participant in most related work \cite{10.1145/3290605.3300437, CHA2013350, murata2001extending}, the use of a cube allowed us to assess rotational variations in our experiment. Due to its identifiable orientation and as one of the most basic 3D shapes, the cube presented a suitable choice to assess both translational and rotational tasks. Regarding rotation, we instructed participants to match the sides of the cube with that of the cube target, as parallel as possible. While using a cube introduces in essence four ``correct'' rotations and limits to some extent the range of rotations one can investigate (e.g. to a maximum of 45 degrees), it still represents the dominant and most widely used 3D shape in current work \cite{9076603}. Our approach is influenced by \cite{8998368, doi:10.1177/0018720810366560}, which included rotational tasks, but were limited to 2D movements only following straight lines without directions and inclinations. The targets were arranged in spherical coordinates with the object at the centre. \autoref{2DTaskTranslationRotation} and \autoref{figure:teaserImage} illustrate movements in 2D and full 3D space respectively.

\begin{figure}
    \centering
    \begin{subfigure}{.32\textwidth}
      \centering
      \includegraphics[width=1\linewidth]{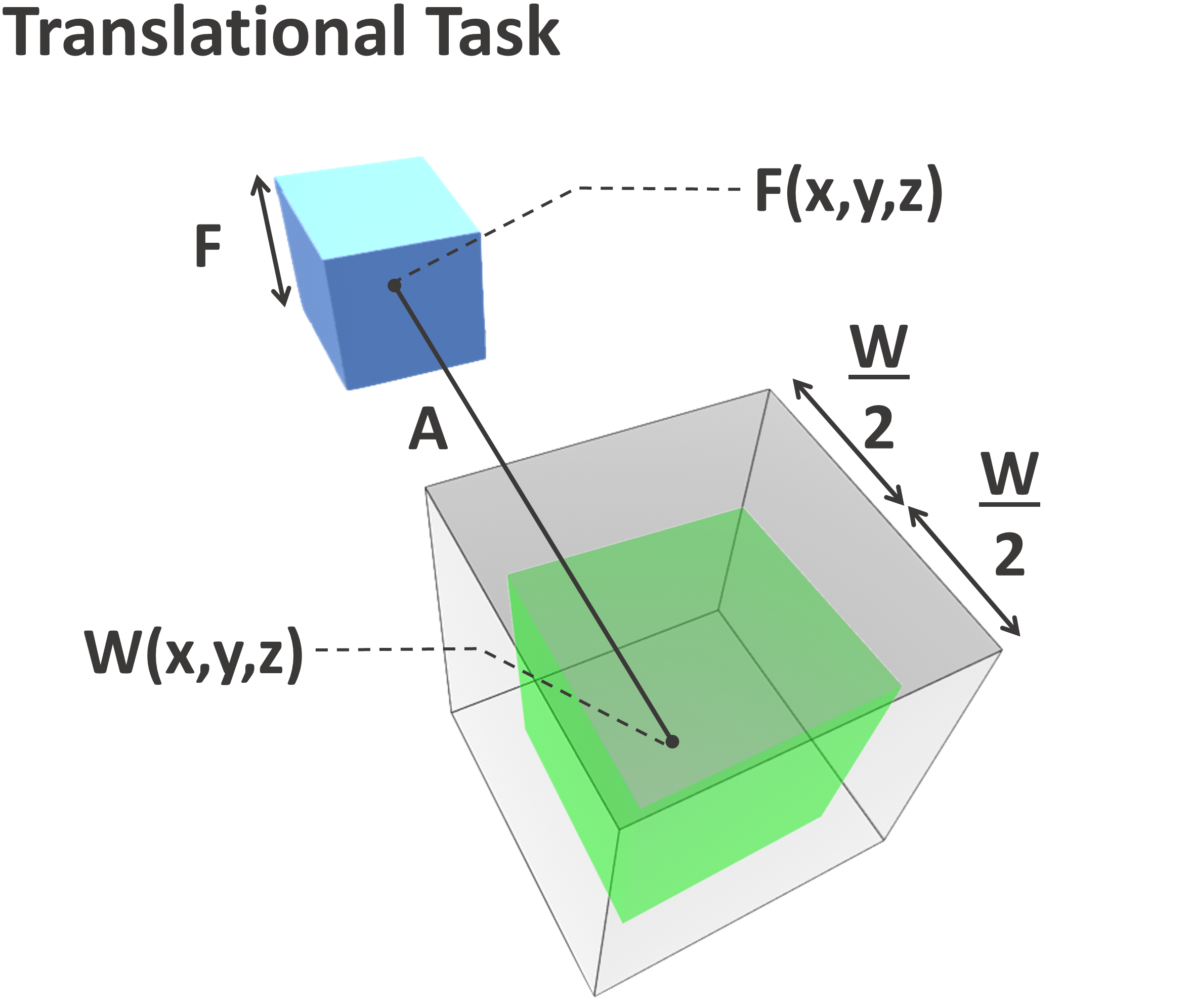}
    \end{subfigure}%
    \begin{subfigure}{.32\textwidth}
      \centering
      \includegraphics[width=1\linewidth]{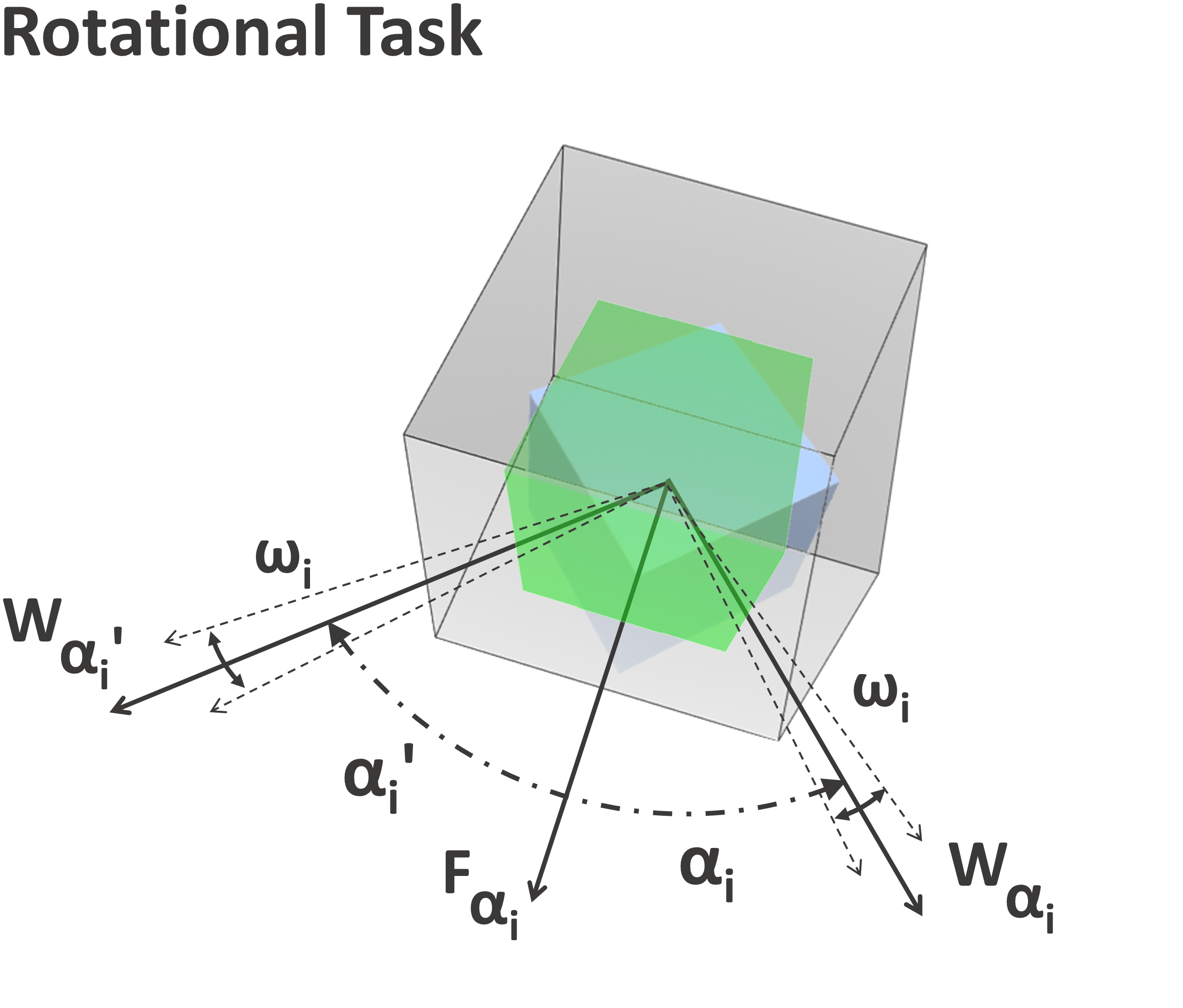}
    \end{subfigure}%
    \begin{subfigure}{.32\textwidth}
      \centering
      \includegraphics[width=1\linewidth]{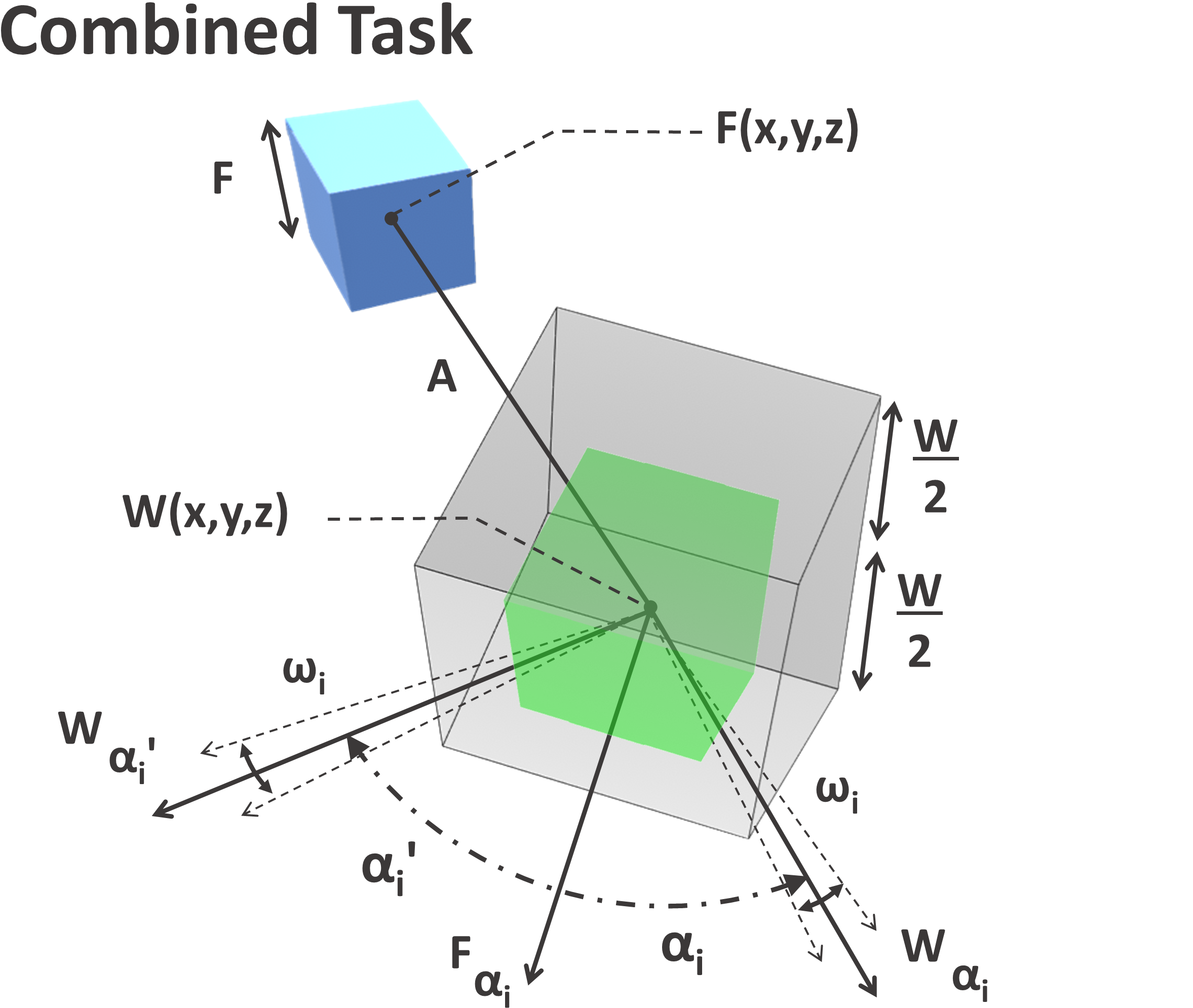}
    \end{subfigure}
    \bigskip%
    \vspace{-0.5cm}
    \begin{subfigure}{\columnwidth}%
        \begin{table}[H]\centering
            \begin{scriptsize}
                \begin{tabular}{p{5.325cm}p{5.325cm}p{5.325cm}} \toprule
                \textbf{Translational Tasks} & \textbf{Rotational Tasks} & \textbf{Combined Tasks} \\\midrule
                $T = \begin{cases}
                             1,  & A \leqslant \frac{W}{2} \\
                             0,  & A > \frac{W}{2}
                       \end{cases}$
                &
                $R = \begin{cases}
                             1,  & \alpha_{i} \leqslant \omega_{i} \\
                             0,  & \alpha_{i} > \omega_{i}
                       \end{cases}$, $i:\mathbb{R}^3$
                &
                $C = T \cdot R$
                \\
                \midrule
                \multicolumn{3}{p{16cm}}{\textbf{Notations for Translational Tasks:}
                \(A\) represents the distance between two 3D points, hence $A = d(W_{x,y,z},F_{x,y,z}) = \sqrt{\sum_{i=1}^{3}\left ( x_{i} - y_{i} \right )^2}$.}\\
                \multicolumn{3}{p{16cm}}{\textbf{Notations for Rotational Tasks:}
                $\alpha_{i} = \left | W_{\alpha_{i}} - F_{\alpha_{i}} \right |$. $W_{\alpha_{i}}$ and $F_{\alpha_{i}}$ represent the rotation of the target and the object in all axes (\(i\)) respectively. $\alpha'_{i} = \alpha_{i}$.}\\
                \multicolumn{3}{p{16cm}}{\textbf{Notations for Combined Tasks:} Both translational and rotational requirements need to be met to be classified as success e.g. \(T=1\) and \(R=1\).}\\
                \bottomrule
                \end{tabular}
            \end{scriptsize}
        \end{table}
    \end{subfigure}
    \caption{From left to right, the figure represents the progression requirements for Translational, Rotational and Combined tasks. The green cube volume illustrates successful placements. The table represents the task successes (1) or errors (0).}
\label{fig:tasksuccessimagemaths}
\end{figure}

The interactive object manipulated by users was presented as a solid blue cube and the target location as a transparent cubic volume. When intersected and translational and/or rotational requirements met, depending on the task type and difficulty, the transparent target would turn green, indicating the success of the task and progressing to the next, as shown in \autoref{figure:ManipulationType} and \autoref{figure:PointingType}. In the cases of pure translational tasks, a 50\% overlap with the target was considered a success as with Fitts' original experiment. For rotational tasks, the target rotation, $\alpha$, needed to be matched within a certain rotation tolerance $\omega$, e.g. the overlap of object-to-target had to be matched in all axes. In \autoref{fig:tasksuccessimagemaths} we visually depict and describe the mathematical equations that needed to be satisfied in order for each type of task to be classified either as a success or an error.

\subsection{\textbf{Pointing vs Manipulation}}
In addition to varying all possible spatial variables in 3D object interaction, we also investigated pointing and manipulation tasks, which are the most common types of interactions in both VEs \cite{10.1145/3290605.3300437} and teleoperated simulation environments \cite{9076603, doi:10.1177/154193128703100723}. Pointing tasks were investigated as still being the dominant interaction type in 3D user interfaces and it closely resembles Fitts' original experiment. Furthermore, pointing tasks are a natural extension of peg insertion tasks, commonly seen in teleoperation \cite{doi:10.1177/154193128703100723}. To cover the most critical limitation in current literature in using Fitts' law for robotics, we also studied manipulation tasks e.g. using grasping, as these are the dominant types of object interaction in teleoperation \cite{10.1145/3411763.3443442}.

For pointing tasks, we used the index pad finger of the virtual hand, to ``attach" the cube to the participant's hand, allowing users to move their hand and the cube to the target location. The object would attach or collide during the intersection between the index finger and the object, and it would match the position of the pad index finger but retain its original orientation. On the other hand, for the manipulation tasks, realistic physics and friction forces were simulated, where participants had to grasp and transport the object to the target location by simulated contact forces. Gravity was enabled for manipulation and a table was simulated by a collision plane. \autoref{figure:PointingType} and \autoref{figure:ManipulationType} visually depict these two types of object interactions.

\section{ANALYSIS AND PROCEDURE}
All results have been tested for significance (95\% CI) via the use of a Repeated-Measures ANOVA (RM-ANOVA). A Shapiro-Wilk Test was used to verify the normality of the data prior to the RM-ANOVA. Maulchy's Test of sphericity was used to test sphericity and in cases of violation, a Greenhouse-Geisser Correction (GGC) was used to account for the violation and correct the degrees of freedom assuming $\epsilon$ $<$0.75. For non-parametric data violating normality, an Aligned-Rank Transform (ART) \cite{Wobbrock:2011:ART:1978942.1978963}, was used to allow the use of the RM-ANOVA on the ranked data. A step-wise linear regression determined the effect of each variable for conceiving our metric equation and the effect on the predictability of \(MT\). The independent spatial variables were either fitted using the criteria of the probability of $F<.05$ to enter, or $F>.10$ to remove. Finally, we used linear regression analysis ($r^2$) to analyze and compare our proposed model with existing work. Hereinafter, the significance levels are: *$p<$.05, **$p<$.01, ***$p<$.001, and n.s standing for ``not significant''.

\subsection{\textbf{Participants}}
A total number of 20 participants ($N=20$) were recruited in this study (4 females and 16 males), with ages ranging from 19 to 46 ($\mu=27.35$, $\sigma=5.43$). The selection criteria we set during recruitment was that each participant was (i) right-handed, (ii) had healthy hand control with (iii) normal/corrected vision and (iv) was familiar with either video games or VR. Participants that did not meet all of these criteria were excluded from the experiment. Participants were asked to find a balance between minimizing errors and selecting the targets as quickly as they could during target selection and placing.

\subsection{\textbf{Pre-Exposure and Approach}}
Prior to the commencement of the experiment, participants were briefed, gave formal written consent and their Individual Interpupillary Distance (IPD) was measured for the VRHMD. Furthermore, acclimatization to the simulation environment was allowed for all participants prior to the experiment, via a set of 96 training exercises. These covered both pointing and manipulation task types, e.g. 48 for each type. Furthermore, both translational and rotational movements were included in these exercises, as to cover the entirety of all spatial complexities later investigated. The training exercises were specific to the acclimatization procedure and independent of those in the experiment. We also randomized the order of all experiments for all participants to counterbalance potential acclimatization or task adaption. A grand total of 39,040 trials were recorded.

\subsection{\textbf{Experiment Procedure}}
To investigate the applicability of our method in full 3D space, we conducted four increasingly more complex experiments as shown in \autoref{table:StudyVariations}. The breakdown in experiments allowed us to compare existing approaches in increasingly more complex spatial settings. All four experiments included two additional and different types of interactions: pointing and manipulation. Throughout, positional tracking was disabled but head tracking was allowed. Participants were nonetheless asked to retain their body pose without moving their torso. To limit potential tiredness, a resting break of 15 seconds took place for every 15 tasks, where users could rest their hands. To control the duration of all experiments, a maximum number of 15 (pointing) and 20 (manipulation) seconds were given for each task, after which the task ended and was considered an error.

Finally, as manipulation inherits multiple different grasping techniques, generally argued to be six distinctive grasping types according to the Southampton Hand Assessment Procedure (SHAP), we instructed participants to use precision/tip grasping for all manipulation tasks, as shown in \autoref{figure:hand_retargeting_simulation_environment} and \autoref{figure:ManipulationType}. Error trials were excluded from the analysis. All of the aforementioned settings described in this section remained constant for all experiments.

\begin{table}\centering
\begin{scriptsize}
\setlength\tabcolsep{1.5pt}
\begin{tabular}{llrrrrrr} \toprule
 \multicolumn{2}{l}{\multirow{2}{*}{\textbf{Variables}}} & \multicolumn{4}{c}{\textbf{Variations Investigated}} \\ \cmidrule(lr){3-6}
 & & {Experiment 1} & {Experiment 2} & {Experiment 3} & {Experiment 4} \\ \midrule
$F$ & \begin{tabular}[c]{@{}l@{}}\tiny Object Size \end{tabular} & (3, 4, 5) & (5) & (4, 5) & (4)\\
$W$ & \begin{tabular}[c]{@{}l@{}}\tiny Target Width \end{tabular} & (5, 7.5, 10, 12.5) & (5, 10) & (5, 10) & (4, 8)\\
$A$ & \begin{tabular}[c]{@{}l@{}}\tiny Target Separation \end{tabular} & (12, 24, 36, 48) & (12, 24) & (0) & (12, 24)\\ 
$\phi$ & \begin{tabular}[c]{@{}l@{}}\tiny Direction Angle \end{tabular} & (90) & (0, 90, 180, 270) & (0) & (0, 90)\\ 
$\theta$ & \begin{tabular}[c]{@{}l@{}}\tiny Inclination Angle \end{tabular} & (0) & (15, 30, 45) & (0) & (15, 30)\\ 
$\alpha$ & \begin{tabular}[c]{@{}l@{}}\tiny Angular Distance \end{tabular} & (0) & (0) & (15, 30, 45) & (30, 45)\\ 
$\omega$ & \begin{tabular}[c]{@{}l@{}}\tiny Angular Tolerance \end{tabular} & (0) & (0) & (2.5, 5, 7.5, 10) & (7.5,15)\\ \midrule
\multicolumn{2}{l}{\tiny Variations x (Reps.)} &\tiny 48x(5)=240 &\tiny 48x(5)=240 &\tiny 48x(5)=240 &\tiny 64x(4)=256 \\
\bottomrule
\end{tabular}
\end{scriptsize}
\caption{Parameter settings of all spatial variations.}
\label{table:StudyVariations}
\vspace{-5mm}
\end{table}

\section{EXPERIMENTS}
Here we present all experiments (E1 to E4) with increasing spatial complexity, each has their results interpreted and analyzed for both pointing and manipulation. \autoref{table:StudyVariations} illustrates all the variations of the investigated variables. \autoref{ResultsAllExperimentsVariablesMT} visually depicts the relationship between these variables and \(MT\); and \autoref{fig:StatisticalResultsVariableContribution} summarizes the statistical results and the significance levels. Finally in \autoref{table:ComparedMethodsFitted} and \autoref{fig:RegressionResults}, we summarize and compare all models with their respective $r^2$ values.

\subsection{\textbf{Experiment 1: Purely Translational Tasks}}
In this experiment, we were primarily interested in approximating Fitts' original experiment, except in 3D. Hence, we investigated the effects of purely translational tasks on the variables of object size \(F\), target width \(W\) and target separation \(A\). Movements were along one line only, with no directions or inclinations as to keep the 3D task simple.

\subsubsection{\textbf{Design}}
This study used a 3x4x4 within-subjects design, and the independent variables were: three object sizes (\(F\) = 3, 4 and 5cm), four target sizes (\(W\) = 5, 7.5, 10, 12.5cm) and four target separations (\(A\) = 12, 24, 36, 48cm). The dependent variable was \(MT\). Among 48 types of tasks, each had 5 repetitions for both pointing and manipulation, hence a total of 480 trials. With 20 participants, a total of 9600 trials were recorded.

\subsubsection{\textbf{Pointing Results}}
The data was not normally distributed, as the Shapiro-Wilk Test yielded $p=.029$, and thus an ART was used prior to the RM-ANOVA to allow the analysis on the ranked data. The mean \(MT\) for pointing tasks was 1.63 $\pm$ 0.40s. Four trials out of 4800 were excluded from the analysis due to errors. Our results showed that an increase in object size \(F\) and target width \(W\) significantly decreased \(MT\), ($p<.01$) and ($p<.001$) respectively. Compared to $F$ and $W$, the target separation $A$ had the biggest effect on \(MT\), with an almost linearly increasing correlation ($p<.001$).

\subsubsection{\textbf{Manipulation Results}}
The data was normally distributed ($p=.229$) and the mean \(MT\) was 2.13 $\pm$ 0.45s. In total 44 trials out of 4800, i.e. 0.9\% of the data, were excluded from the analysis due to errors. \(MT\) significantly decreased as object size $F$ and target width $W$ increased, ($p<.001$) and ($p<.001$) respectively. The target separation $A$ had again the biggest effect, almost linearly increasing with \(MT\) ($p<.001$).

\begin{table*}\centering
\begin{scriptsize}
\setlength\tabcolsep{4.5pt}
\begin{tabular}{llcccccccccccccc} \toprule
 \multicolumn{2}{l}{\multirow{2}{*}{\textbf{}}} &
 \multicolumn{7}{c}{{\textbf{Pointing}}} &
 \multicolumn{7}{c}{{\textbf{Manipulation}}} \\ \cmidrule(lr){3-9} \cmidrule(lr){10-16}
 && \multicolumn{3}{c}{{\textit{\scriptsize Sphericity Test \normalsize}}} & \multicolumn{3}{c}{{\textit{\scriptsize RM-ANOVA Test \normalsize}}} && \multicolumn{3}{c}{{\textit{\scriptsize Sphericity Test \normalsize}}} & \multicolumn{3}{c}{{\textit{\scriptsize RM-ANOVA Test \normalsize}}} \\ \cmidrule(lr){3-5} \cmidrule(lr){6-8} \cmidrule(lr){10-12} \cmidrule(lr){13-15}
 && $\chi^2$ & $p$ & $\epsilon$ & $F$-Test & $\eta_{p}^2$ & $p$ & $r^2$ (\%) & $\chi^2$ & $p$ & $\epsilon$ & $F$-Test & $\eta_{p}^2$ & $p$ & $r^2$ (\%)\\ \midrule
\addlinespace[-0.0025cm] 
\rowcolor{Gray} \multicolumn{16}{c}{{\textbf{Experiment 1}}} \\ \hline 
$F$ && $\chi^2(2)$=2.468 & n.s & n/a & F(2,30)=9.441 & 0.386 & ** & 0.5\% & $\chi^2(2)$=3.314 & n.s & n/a & F(2,30)=14.468 & 0.491 & *** & 3.1\% \\
$W$ && $\chi^2(5)$=6.992 & n.s & n/a & F(3,33)=102.522 & 0.903 & *** & 10.2\% & $\chi^2(5)$=5.409 & n.s & n/a & F(3,33)=19.162 & 0.635 & *** & 5.3\% \\
$A$ && $\chi^2(5)$=8.360 & n.s & n/a & F(3,33)=602.976 & 0.982 & *** & 80.4\% & $\chi^2(5)$=5.968 & n.s & n/a & F(3,33)=197.465 & 0.947 & *** & 84.7\% \\ 

\hline
\rowcolor{Gray} \multicolumn{16}{c}{{\textbf{Experiment 2}}} \\ \hline 
$W$ && n/a & n/a & n/a & F(1,23)=287.752 & 0.925 & *** & 40.0\% & n/a & n/a & n/a & F(1,23)=51.431 & 0.691 & *** & 35.6\% \\
$A$ && n/a & n/a & n/a & F(1,23)=190.993 & 0.893 & *** & 43.7\% & n/a & n/a & n/a & F(1,23)=33.986 & 0.596 & *** & 23.0\% \\
$\phi$ && $\chi^2(5)$=12.609 & * & 0.675 & F(2.025,22.270)=6.918 & 0.386 & * & 3.0\% & $\chi^2(5)$=3.722 & n.s & n/a & F(3,33)=0.851 & 0.072 & n.s & 1.0\%\\
$\theta$ && $\chi^2(2)$=4.346 & n.s & n/a & F(2,30)=7.160 & 0.323 & * & 1.3\% & $\chi^2(2)$=2.210 & n.s & n/a & F(2,30)=5.535 & 0.270 & * & 5.4\% \\ 

\hline 
\rowcolor{Gray} \multicolumn{16}{c}{{\textbf{Experiment 3}}} \\ \hline 
$F$ && n/a & n/a & n/a & F(1,23)=1.551 & 0.063 & n.s & 0.0\% & n/a & n/a & n/a & F(1,23)=1.195 & 0.049 & n.s & 0.5\%\\
$W$ && n/a & n/a & n/a & F(1,23)=0.299 & 0.013 & n.s & 0.0\% & n/a & n/a & n/a & F(1,23)=40.805 & 0.640 & *** & 6.4\%\\
$\alpha$ && $\chi^2(2)$=0.781 & n.s & n/a & F(2,30)=23.639 & 0.612 & *** & 1.8\% & $\chi^2(2)$=3.589 & n.s & n/a & F(2,30)=12.915 & 0.463 & *** & 2.9\%\\
$\omega$ && $\chi^2(5)$=11.386 & * & 0.639 & F(1.918,21.093)=184.323 & 0.944 & *** & 81.8\% & $\chi^2(5)$=4.454 & n.s & n/a & F(3,33)=89.481 & 0.891 & *** & 73.3\%\\ 

\hline 
\rowcolor{Gray} \multicolumn{16}{c}{{\textbf{Experiment 4}}} \\ \hline 
$W$ && n/a & n/a & n/a & F(1,31)$<$0.001 & 0.0 & n.s & 0.0\% & n/a & n/a & n/a & F(1,31)=58.815 & 0.655 & *** & 13.1\%\\
$A$ && n/a & n/a & n/a & F(1,31)=51.296 & 0.623 & *** & 8.5\% & n/a & n/a & n/a & F(1,31)=48.664 & 0.611 & *** & 19.2\%\\
$\phi$ && n/a & n/a & n/a & F(1,31)=2.278 & 0.680 & n.s & 0.7\% & n/a & n/a & n/a & F(1,31)=0.205 & 0.007 & n.s & 0.1\% \\
$\theta$ && n/a & n/a & n/a & F(1,31)=1.718 & 0.053 & n.s & 0.2\% & n/a & n/a & n/a & F(1,31)=2.990 & 0.088 & n.s & 0.7\% \\
$\alpha$ && n/a & n/a & n/a & F(1,31)=10.548 & 0.254 & ** & 1.2\% & n/a & n/a & n/a & F(1,31)=10.193 & 0.247 & ** & 4.0\%\\
$\omega$ && n/a & n/a & n/a & F(1,31)=394.944 & 0.927 & *** & 77.3\% & n/a & n/a & n/a & F(1,31)=122.753 & 0.798 & *** & 44.3\%\\
\bottomrule
\end{tabular}
\end{scriptsize}
\caption{Results of all variables and their influence towards \(MT\), across all experiments. Percentages from $r^2$ (\%) indicate the weight or contribution of each variable towards \(MT\) and were obtained via step-wise linear regression.}
\label{fig:StatisticalResultsVariableContribution}
\vspace{-5mm}
\end{table*}

\subsubsection{\textbf{Remarks}}
From our analysis, we observed that all included variables had a significant effect on \(MT\) for both pointing and manipulation. Fitting the models of Fitts', Shannon's and Welford's using the data, the results showed a high fitting for pointing, but a slightly less fitting for manipulation. Only Hoffmann's model showed a significantly better fitting for manipulation, at an approximately 10\% increase from the rest, which is likely due to the incorporation of the object size $F$ in the formulation. Murata \& Iwase' and Cha \& Myung's were excluded from this analysis as they yielded the same results as their based extensions, since directions and inclinations were not assessed in E1.

\subsection{\textbf{Experiment 2: Translational Tasks with Directions and Inclinations}}
E2 is an extension of E1 by adding directions and inclinations, which have significant effects according to \cite{CHA2013350, murata2001extending}.

\subsubsection{\textbf{Design}}
The study used a 2x2x4x3 within-subjects design, and the independent variables were: two target sizes (\(W\) = 5 and 10cm), two target separations (\(A\) = 12 and 24cm), four direction angles ($\phi$ = 0$^{\circ}$, 90$^{\circ}$, 180$^{\circ}$ and 270$^{\circ}$) and three inclination angles ($\theta$ = 15$^{\circ}$, 30$^{\circ}$ and 45$^{\circ}$). The dependent variable was \(MT\). A total of 9600 trials were recorded.

\subsubsection{\textbf{Pointing Results}}
The data was normally distributed ($p=.077$) and the mean \(MT\) for pointing tasks was 1.47 $\pm$ 0.23s. A total of 3 trials out of 4800 were excluded from the analysis due to errors. The target width $W$ significantly decreased \(MT\) with higher dimensions ($p<.001$), while high values of target separation $A$ significantly increased \(MT\) ($p<.001$). Directional angles $\phi$ significantly affected \(MT\) as well ($p<.05$), with a slight sinusoidal relationship with \(MT\) and revealing that front-to-backward movements (0$^{\circ}$, 180$^{\circ}$) take slightly longer time than left-to-right ones (90$^{\circ}$, 270$^{\circ}$), in line with \cite{10.1145/3290605.3300437}. Finally, higher degrees of inclinations $\theta$ translated in longer \(MT\), ($p<.05$), matching the findings of \cite{CHA2013350}.

\subsubsection{\textbf{Manipulation Results}}
The data was normally distributed ($p=.083$) and the mean \(MT\) for manipulation tasks was 2.11 $\pm$ 0.25s. A total of 60 trials out of 4800, 1.25\% of the data, were excluded from the analysis due to error. \(MT\) significantly decreased as target width $W$ increased ($p<.001$). Consistent with E1, an increase in target separation $A$ significantly increased \(MT\)  ($p<.001$). Contrary to pointing tasks, directional angles $\phi$ did not have an apparent effect on \(MT\) ($p=.476$). Finally, increased incline angles $\theta$ significantly affected \(MT\) ($p<.05$) as in \cite{CHA2013350}.

\subsubsection{\textbf{Remarks}}
All variables had significant effects on \(MT\), except directional angles $\phi$ in manipulation. Contrary to E1, we observed that Fitts', Shannon's and Welford's formulation had a relatively higher fitting both in pointing and manipulation than Hoffmann's model, by approximately 5\%. This is confirmed with Cha \& Myung's model, which is based on Hoffmann's, having a very similar fitting for both pointing and manipulation. Only Murata \& Iwase's directional model presented a marginally higher fitting than the rest in pointing and manipulation tasks.

\subsection{\textbf{Experiment 3: Purely Rotational Tasks}}
To this point, only translational tasks were investigated. Yet, from the identified literature and intuition, rotation is an inseparable and fundamental part \cite{10.1145/3411763.3443442, 8998368, doi:10.1177/0018720810366560}. Hence, here translation is eliminated as both the object and the target will be overlapping in the centre, with only their rotation differing. Consequently, this task is an almost natural equivalent of Fitts' original experiment, only for rotation.

\subsubsection{\textbf{Design}}
The study used a 2x2x3x4 within-subjects design, in which the independent variables were: two object sizes (\(F\) = 4 and 5cm), two target sizes (\(W\) = 5 and 10cm), three target rotations ($\alpha$ = 15$^{\circ}$, 30$^{\circ}$ and 45$^{\circ}$) and four rotational tolerances ($\omega$ = 2.5$^{\circ}$, 5$^{\circ}$, 7.5$^{\circ}$ and 10$^{\circ}$). The dependent variables was \(MT\). 9600 trials were recorded.

\subsubsection{\textbf{Pointing Results}}
The data deviated from normality ($p<.001$) and an ART was applied, and the mean \(MT\) for pointing tasks was 3.37 $\pm$ 1.51s. A total of 225 trials out of 4800, 4.6\% of the data, were excluded from the analysis due to error. An increase in both object size $F$ and target width $W$ showed a slight decrease towards \(MT\), but not at a significant level, ($p=.226$) and ($p=.590$) respectively. However, we observed a significant correlation between the rotational separation $\alpha$ and \(MT\) ($p<.001$). The largest effect was observed with the rotational tolerance $\omega$, showing an almost inverse exponential relationship with \(MT\) ($p<.001$). We can infer that the effect of $\omega$ was so significant that it had almost the same inverse effect on \(MT\) as the equivalent of $W$ and $F$ for translational tasks, though at a higher degree.

\subsubsection{\textbf{Manipulation Results}}
The data violated normality ($p<.01$) and an ART was thus used. The mean \(MT\) for manipulation tasks was 2.79 $\pm$ 0.67s. A total of 64 trials out of 4800, 1.3\% of the data, were excluded due to error. In line with pointing, no significant effect was observed on \(MT\) with the object size $F$ ($p=.286$). In contrast to pointing, an increase in target width $W$ did significantly decrease \(MT\) ($p<.001$). Increasing rotational separation $\alpha$ also increased \(MT\) significantly ($p<.001$). Finally, as with pointing, increasing rotational tolerance $\omega$ significantly decreased \(MT\) ($p<.001$) in a non-linear fashion. With the exception of target width $W$, we observed the same relationship of the tested variables on \(MT\) for both pointing and manipulation. Rotational tolerance $\omega$ predominately affected \(MT\) in a purely rotational setting. This was further investigated in the final experiment E4 to increase our understanding of the underlying factors.

\begin{figure*}
\centering
  \includegraphics[width=1.0\textwidth]{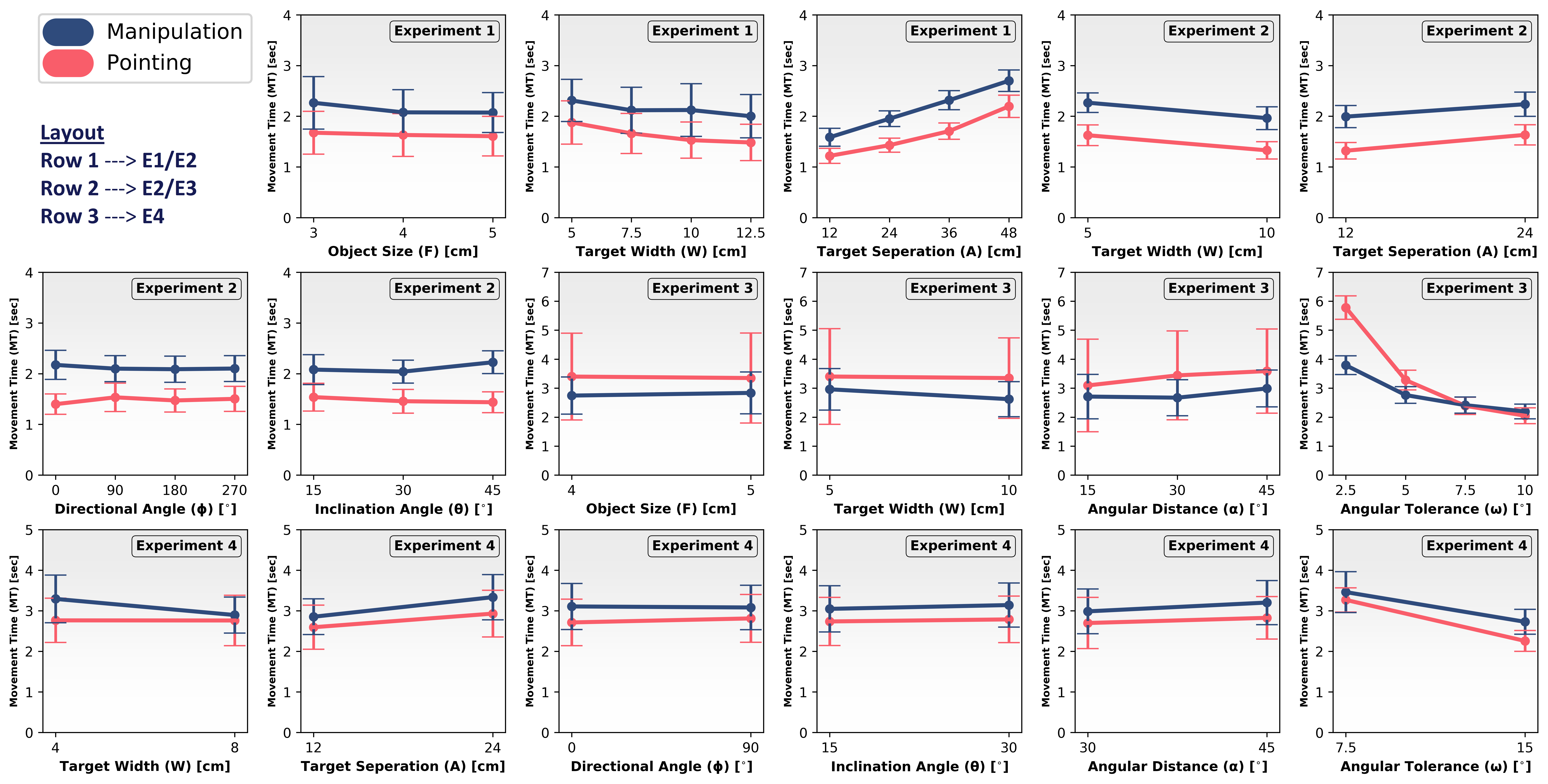}
  \vspace{-5mm}
 \caption{Results of all tested variables in all experiments and the relationship with \(MT\) for both pointing (red) and manipulation (blue). Figure sequence is from left to right and top to bottom. Bars represent the standard deviation.}~\label{ResultsAllExperimentsVariablesMT}
  \vspace{-5mm}
\end{figure*}

\begin{figure*}
\centering
  \includegraphics[width=1.0\textwidth]{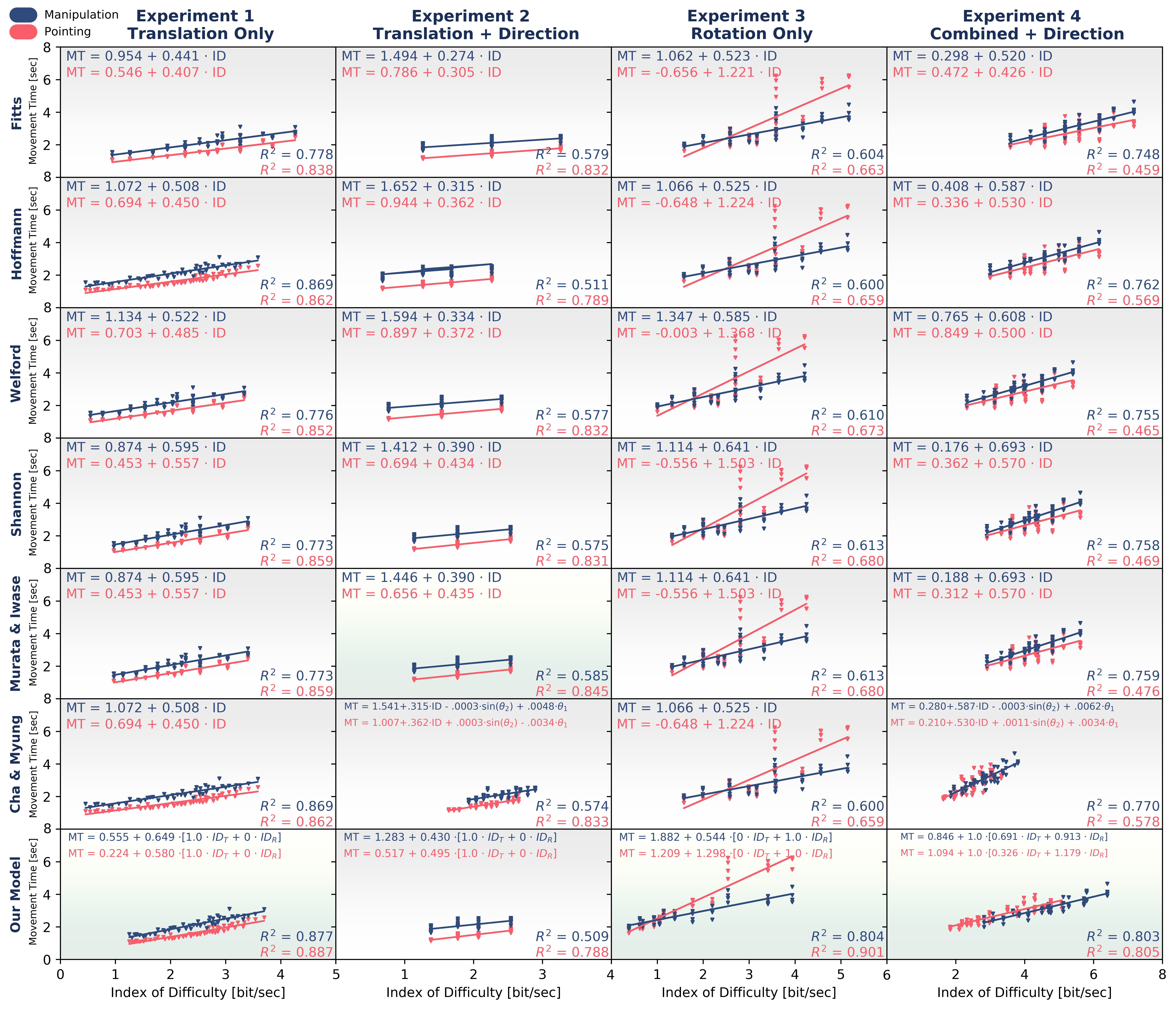}
  \vspace{-7.5mm}
 \caption{Regression plots of all models across E1 to E4, depicting line equations and $r^2$. Green boxes represent best model.}~\label{fig:RegressionResults}
  \vspace{-7.5mm}
\end{figure*}

\subsubsection{\textbf{Remarks}}
For this experiment, we fit all existing equations to accommodate the rotational nature of this experiment, as detailed in \autoref{table:ComparedMethodsFitted}. By doing so, we are able to conduct and apply these methods to the rotational nature of this experiment, and also investigate any significant effects for comparison. We further extend the work of \cite{doi:10.1177/0018720810366560} and cover all known extensions. Overall, regarding variables influencing \(MT\), rotational variables $\alpha$ and $\omega$ were significant in both cases, although some inconsistencies were observed between pointing and manipulation. Noticeably, none of the adjusted models fitted this rotational task well, where all formulations were below $r^2 \leq 0.680$ for both pointing and manipulation. Contrary to the translational experiments (E1 \& E2), we see that these models did not fit the data very well in rotational tasks.

\subsection{\textbf{Experiment 4: Combined Translational and Rotational Tasks with Directions and Inclinations}}
For E4, we combined all possible variations from E1, E2 and E3, constituting this experiment a fully combined movement task. We combine both translational and rotational task variations with varying directional and inclination gains to describe a full 3D task. To limit the number of variations, we restricted the directional movements $\phi$ towards the front and right side direction (0$^{\circ}$, 90$^{\circ}$), still investigating view and lateral directional influences. We also kept the object size fixed at \(F\) = 4cm. These design decisions were made to limit the number of tasks, allowing us to control our already exhaustive experiment.

\subsubsection{\textbf{Design}}
For the final experiment, a 2x2x2x2x2x2 within-subjects design was used, in which the independent variables were: two target sizes (\(W\) = 4 and 8cm), two target separations (\(A\) = 12 and 24cm), two direction angles ($\phi$ = 0$^{\circ}$ and 90$^{\circ}$), two inclination angles ($\theta$ = 15$^{\circ}$ and 30$^{\circ}$), two target rotations ($\alpha$ = 30$^{\circ}$ and 45$^{\circ}$) and two rotational tolerances ($\omega$ = 7.5$^{\circ}$ and 15$^{\circ}$). The dependent variable was \(MT\). Representing 64 tasks and with 4 repetitions, we conducted 256 trials for both pointing and manipulation tasks, i.e. 512 trials per participant resulting in 10240 trials in total.

\subsubsection{\textbf{Pointing Results}}
An ART was applied since normality was violated ($p=.016$). The mean \(MT\) for pointing tasks was 2.71 $\pm$ 0.55s. A total of 120 error trials out of 4800, 2.5\% of the data, were excluded from the analysis. Sphericity was met in all cases since all variations of the tested variables were two levels. There was no significant effect of target width $W$ on \(MT\), ($p>.05$). On the other hand, the target separation $A$, consistent with E1 and E2, significantly increased \(MT\) with higher separation values ($p<.001$). Neither directional arrangements $\phi$ nor inclination angles $\theta$ had a significant effect on \(MT\), ($p=.141$) and ($p=.200$). As for rotation, $\alpha$ significantly affected \(MT\) ($p<.01$) and $\omega$ had the highest influence on \(MT\) ($p<.001$), consistent with E3.

\subsubsection{\textbf{Manipulation Results}}
The data was normally distributed ($p=.091$). The mean \(MT\) for manipulation tasks was 3.10 $\pm$ 0.59s. A total of 82 trials out of 4800, 1.7\% of the data, were excluded from the analysis due to error. Sphericity was met in all cases. Contrary to the pointing task, target width $W$ had a significant effect on \(MT\) ($p<.001$). Target separation $A$ significantly increased \(MT\) with higher separation values ($p<.001$). Directional angles $\phi$, as well as inclination angles $\theta$, followed the same trend as with the pointing equivalent, not affecting \(MT\) much ($p=.099$) and ($p=.227$) respectively. Rotational separation $\alpha$ did significantly affect the increase of \(MT\) as the degrees of separation rose ($p<.01$). Rotational tolerance $\omega$ had again the biggest influence on \(MT\), being consistent with the pointing equivalent and E3 ($p<.001$).

\subsubsection{\textbf{Remarks}}
For E4, we added the \(IDs\) of translation and rotation of all existing approaches. The rotation was adjusted as described in E3 and in \autoref{table:ComparedMethodsFitted}. Despite these models in principle being incompatible for combined movements, we added these two \(IDs\) to explore the potential of summing these two separate concepts of motion and to conduct a fair comparison. For example, the combined \(ID\) for Fitts' model would be $ID_{t} =log_2(\frac{2A}{W})$ for translation, plus $ID_{r} =log_2(\frac{2\alpha}{\omega})$ for rotation, with the other formulations following the same fashion. Overall, neither directional $\phi$ nor inclination $\theta$ angles had any significant effect on \(MT\) ($p>0.05$) in pointing or manipulation. For combined movements, adding both \(IDs\) of translation and rotation, Fitts's, Shannon's, Welford's and even Murata \& Iwase's directional formulation proved insufficient when fitted for pointing ($r^2<0.5$), but yielded slightly better results for manipulation ($r^2<0.77$). Hoffmann's had a significantly higher fitting towards pointing with an approximately 10\% increase. Being consistent with E1, this suggests the necessity of adding the object size $F$ in a model. Cha \& Myung's had overall the highest fitting on the data but only marginally over the rest, primarily due to the incorporation of object size \(F\) as seen also in Hoffmann's.

\section{MODEL DERIVATION}
\label{sec:modelDerivation}
In most prior work related with extending Fitts' law, a clear definition of distance is mostly missing \cite{10.1145/1753846.1753867}. For example, the distance from the object to the target ($A$) can either be defined from the object centre to the target centre ($A_{cc}$), the object centre to the target edge ($A_{ce}$) or even the object edge to the target edge ($A_{ee}$). Such a discrepancy highly influences the formula and can be formulated as:
\begin{equation}
\label{eqn:DistanceDefinitionSimple}
\begin{gathered}
A_{n} 
\begin{cases}
A_{1} = A_{cc}, \\
A_{2} = A_{ec} + \frac{W}{2},\\
A_{3} = A_{ee} + \frac{W+F}{2},\\
\end{cases}
d_{n} = \sqrt{\sum_{i=1}^{n}\left ( x_{i} - y_{i} \right )^2}.
\end{gathered}
\end{equation}

From the above, a total of 3 different target separations are defined as $A_{n}$ with $A_{cc}>A_{ec}>A_{ee}$. The Euclidean distance between two points in 3D is represented as $d_{n}$ with $n=3$. At each stage of the experiments, our results show that 3D translational movement follows Fitts' formulation closely. Target separation $A$ and target width $W$ are an integral part of the formulation. However, another variable with a determining effect in translational tasks was the object size $F$, as part of Hoffmann's model but not of Fitts'. Hence, the most appropriate definition of effective target separation would be $A_{3}$ from Eq. \ref{eqn:DistanceDefinitionSimple} which would include both $W$ and $F$. By incorporating this definition into Fitts' model \(ID\) shown in Eq. \ref{eqn:fitts_law_original}, we have:
\begin{equation}
\label{eqn:finalTranslation}
\begin{gathered}
    ID_{t} = log_2 \left ( \frac{2A}{F + W} + 1\right ).\\
\end{gathered}
\end{equation}

As for rotational movements, the graphs in \autoref{ResultsAllExperimentsVariablesMT} show that while $\alpha$ follows an almost linear relationship with \(MT\), and $\omega$ follows an evident non-linear relationship. While $\omega$ follows an almost inverse exponential decay, we believe a $\frac{1}{x^2}$ would be more appropriate, yet more data points would be needed to shed additional light on this discrepancy.

With the results at hand and the indicated discrepancy between translation and rotation as well as using the aforementioned distance definition, we can formulate an improved and more suitable 3D model as:
\begin{equation}
\label{eqn:finalModel}
\begin{aligned}
\text{Final Model: }
\begin{cases}
MT &= a + b [c \cdot ID_{t} + d \cdot ID_{r}], \\
ID_{t} &= log_2 \left ( \frac{2A}{F + W} + 1\right ),\\
ID_{r} &= log_2 \left ( \frac{2\alpha}{\omega^2} + 1 \right ),\\
\end{cases}
\end{aligned}
\end{equation}
where \(a\), \(b\), \(c\) and \(d\) are constants determined through regression. We retain Fitts' original approach, and hence constants \(a\) and \(b\) are the y-intercept and slope respectively. Constants \(c\) and \(d\) represent the contribution of translation (\(ID_{t}\)) and rotation (\(ID_{r}\)) respectively, as these are two separate concepts of motion from our analyzed results. It is worthwhile to point out that the constant \(b\) is superfluous in this case since it multiplies constants \(c\) and \(d\). However, the addition of the \(b\) constant allows the resemblance and familiarity of our model to that of Fitts' original formulation as seen in Eq. \ref{eqn:fitts_law_original}.

This model follows Stoelen \& Akin's proposed formulation \cite{doi:10.1177/0018720810366560}, and in particular the feasibility of adding the \(IDs\) of both translation and rotation into one combined index of difficulty. Yet in the formulation of \cite{doi:10.1177/0018720810366560}, both $ID_{t}$ and $ID_{r}$ were controlled by a single constant \(b\), as $MT = a + b \cdot [ID_{t} + ID_{r}]$. However, the assumption that translation and rotation have an equal contribution (i.e. $50\%$ or 0.5) towards task difficulty (\(ID\)) and by extent \(MT\) is imprecise and limited. 

In particular, our results show that rotation almost predominately affected \(MT\) and was primarily attributed to high rotational accuracy ($\omega$ = 2.5$^{\circ}$ in E3). For very small $\omega$, we observed an almost inverse exponential decay with \(MT\), instead of a linear one. Translation and rotation should hence be perceived as separate terms, and each should be quantified by its own constants. This mitigates the limitation of equal weighting as with \cite{doi:10.1177/0018720810366560}, which is not suitable as shown in our study. In the following section, we analyze the results and compare the proposed model with the other aforementioned versions.

\begin{table*} \centering
\begin{footnotesize}
\setlength\tabcolsep{6.25pt}
\begin{tabular}{lrrrrrrrr} \toprule
 \multirow{3}{*}{\textbf{Model Fit ($\mathit{r^2}$)}} & \multicolumn{2}{c}{\textbf{Experiment 1}} & \multicolumn{2}{c}{\textbf{Experiment 2}} & \multicolumn{2}{c}{\textbf{Experiment 3}} & \multicolumn{2}{c}{\textbf{Experiment 4}}\\ 
 \multirow{2}{*}{\textbf{}} & \multicolumn{2}{c}{\scriptsize(Translation Only)\normalsize} & \multicolumn{2}{c}{\scriptsize(Translation w/ Directions)\normalsize} & \multicolumn{2}{c}{\scriptsize(Rotation Only)\normalsize} & \multicolumn{2}{c}{\scriptsize(Trans. w/ Rot. \& Directions)\normalsize}\\ 
  \cmidrule(lr){2-3} \cmidrule(lr){4-5} \cmidrule(lr){6-7} \cmidrule(lr){8-9}
 & \multicolumn{1}{c}{\scriptsize Pointing \normalsize} & \multicolumn{1}{c}{\scriptsize Manipulation \normalsize} & \multicolumn{1}{c}{\scriptsize Pointing \normalsize} & \multicolumn{1}{c}{\scriptsize Manipulation \normalsize} & \multicolumn{1}{c}{\scriptsize Pointing \normalsize} & \multicolumn{1}{c}{\scriptsize Manipulation \normalsize} & \multicolumn{1}{c}{\scriptsize Pointing \normalsize} & \multicolumn{1}{c}{\scriptsize Manipulation \normalsize}\\  \midrule
 
 Fitts' \cite{fitts1954information} & \gradient{0.838} & \gradient{0.778} & \gradient{0.832} & \gradient{0.579} &  \hspace{-0.35cm} \gradient{0.663} &   \hspace{-0.35cm} \gradient{0.604} &   \hspace{-0.35cm} \gradient{0.459} &   \hspace{-0.35cm} \gradient{0.748}\\
 Hoffmann's \cite{doi:10.1080/00140139508925153} & \gradient{0.862} & \gradient{0.869} & \gradient{0.789} & \gradient{0.511} &  \hspace{-0.35cm} \gradient{0.659} &  \hspace{-0.35cm} \gradient{0.600} &  \hspace{-0.35cm} \gradient{0.569} &  \hspace{-0.35cm} \gradient{0.762}\\
 Welford's \cite{welford1968fundamentals} & \gradient{0.852} & \gradient{0.776} & \gradient{0.832} & \gradient{0.577} &  \hspace{-0.35cm} \gradient{0.673} &  \hspace{-0.35cm} \gradient{0.610} &  \hspace{-0.35cm} \gradient{0.465} &  \hspace{-0.35cm} \gradient{0.755}\\
 Shannon's \cite{10.1207/s15327051hci0701_3} & \gradient{0.859} & \gradient{0.773} &  \gradient{0.831} & \gradient{0.575} &   \hspace{-0.35cm} \gradient{0.680} &   \hspace{-0.35cm} \gradient{0.613} & \hspace{-0.35cm}  \gradient{0.469}  & \hspace{-0.35cm} \gradient{0.758}\\
 Murata \& Iwase \cite{murata2001extending} &  \hspace{-0.35cm} \gradient{0.859} &  \hspace{-0.35cm} \gradient{0.773} & \textbf{\gradient{0.845}} & \textbf{\gradient{0.585}} &  \hspace{-0.35cm} \gradient{0.680} &  \hspace{-0.35cm} \gradient{0.613} &  \hspace{-0.35cm} \gradient{0.476} &  \hspace{-0.35cm} \gradient{0.759}\\
 Cha \& Myung \cite{CHA2013350} &  \hspace{-0.35cm} \gradient{0.862} &  \hspace{-0.35cm} \gradient{0.869} & \gradient{0.833} & \gradient{0.574}  & \hspace{-0.35cm} \gradient{0.659} & \hspace{-0.35cm} \gradient{0.600} &  \hspace{-0.35cm} \gradient{0.578} &  \hspace{-0.35cm} \gradient{0.770}\\
 Our Model & \textbf{\gradient{0.887}} & \textbf{\gradient{0.877}} & \gradient{0.788} & \gradient{0.509} & \textbf{\gradient{0.901}} & \textbf{\gradient{0.804}} & \textbf{\gradient{0.805}} & \textbf{\gradient{0.803}}\\
\bottomrule
\end{tabular}
\end{footnotesize}
\caption{Overview of all methods with the respective $r^2$ fitting. For E3 and E4 where rotation is present, existing models are in theory incompatible. However, to retain consistency and to allow for a fair comparison, as motivated by \cite{doi:10.1177/0018720810366560}, these were adjusted and extended by us to accommodate rotation as well. For example Fitts' $ID_{t} =log_2(\frac{2A}{W})$ would become $ID_{r} =log_2(\frac{2\alpha}{\omega})$ and both \(IDs\) are added for E4. Bold numbers represent the best method in the column group.}
\label{table:ComparedMethodsFitted}
\end{table*}

\section{DISCUSSION}
This study compared the most widely used extensions on Fitt's law through four designed experiments in a simulated teleoperation VR setting. This allowed us to evaluate the applicability of each model with various spatial complexities entailing translational and rotational movements. Each experiment included pointing and manipulation tasks, which are the most widely used types of interactions in collaborative VEs, VR simulators and robot teleoperation \cite{10.1145/3290605.3300437,9076603,doi:10.1177/154193128703100723}.

The study showed that in the most basic form of 3D object movement in a purely translational setting (E1), Fitts' law and its extensions are adequate for both target pointing and manipulation along a one-directional line only. However, when complexity increased by including directional and inclination angles, these models had reduced performance at predicting the results of our experiments, which dropped slightly in pointing tasks but significantly more in manipulation tasks (E2). 

Though Fitts' law has been extended towards rotation, studies so far have limited their findings in 2D space \cite{412031, 8998368, doi:10.1177/0018720810366560}. Furthermore combining rotational and translational movements under one setting in 3D remains largely unexplored. While Kulik et al. \cite{8998368} and Stoelen \& Akin \cite{doi:10.1177/0018720810366560} studied combined movements, these were still limited to 2D space and only following movements across one line. 

These current limitations in the state of the art showed their drawbacks in E3 and E4. All models in a purely rotational setting (E3) were insufficient when extended towards pointing and further aggravated during manipulation. However, the most important and critical experiment was E4, where we combined translational and rotational movements with varying spatial arrangements under one setting, rendering the last experiment a full 3D task. In E4, where we effectively investigated the most complex spatial settings in 3D, all compared models proved insufficient for both pointing and manipulation. These observations led us to the proposal of a new metric to overcome these limitations and the derived model outperformed other extensions in E1, E3 and most importantly in E4, as shown in \autoref{table:ComparedMethodsFitted} and \autoref{fig:RegressionResults}.

Hence, our metric could be used to assess human movement in teleoperation \cite{doi:10.1177/154193128703100723} especially with the use of MR technologies \cite{8283715}. For example, teleoperation studies attempting to evaluate the use of new sensory feedback or techniques, such as bilateral operation with haptic feedback to complete 3D tasks \cite{babarahmati2020robust}, could greatly benefit from a 3D metric on how it affects human performance. We support this from our results, as our formulation can capture the complex spatial settings in 3D space associated with such scenarios. Moreover, as our formulation is based on Fitts' law and hence combines spatial and time-based metrics under a single model, further studies on HCI/HRI could benefit in terms of standardization \cite{10.1145/3411763.3443442, 10.1145/1121241.1121249}.

\begin{table*}\centering
\begin{scriptsize}
\begin{tabular}{p{5mm}p{16.25cm}} \toprule
 \multicolumn{2}{l}{\textbf{Research Implications and Findings}}\\ \midrule
\rowcolor{Gray} \multirow{1}{*}{\textbf{R.1}} & Fitts' law can be extended towards 3D but does not adequately explain all spatial settings, and requires a clear definition of the distances associated in 3D.\\
\multirow{1}{*}{\textbf{R.2}} & Translation and rotation should be classified as separate concepts and quantities. Our analysis showed when rotation was present, it predominantly influenced \(MT\) over translation, by an almost $\frac{2}{3}$. Hence, a model has been built to account for these two separate concepts of motion by incorporating a constant for each, as these do not have an equal weight towards \(MT\) as observed by our experimental results. \\
\rowcolor{Gray} \multirow{1}{*}{\textbf{R.3}} & Pointing and manipulation are perceived as inherently two very different types of tasks, but do follow and can be modelled by Fitts' law even in 3D space. \\
\multirow{1}{*}{\textbf{R.4}} & Object size does matter. During pointing but especially in manipulation this holds true. For the latter, the object size is an integral part of the metric. This is achieved by adding the object size $F$ into the equation as suggested by Hoffmann \cite{doi:10.1080/00140139508925153} and confirmed in our model extension. \\
\rowcolor{Gray} \multirow{1}{*}{\textbf{R.5}} & Rotation tolerance $\omega$ or simply the rotational accuracy, significantly affects \(MT\). This is particularly the case for very low rotational tolerances ($2.5^{\circ}$). Hence, strict rotational accuracy requirements during object placement (i.e. teleoperation) are likely to significantly increase operator completion timings.\\
\midrule
\multicolumn{2}{l}{\textbf{Recommendations for Task Design in Pointing and Manipulation Tasks in VR and Teleoperation}} \\
\midrule
\rowcolor{Gray} \multirow{1}{*}{\textbf{RC1}} & Avoid having ``strict" spatial requirements towards accuracy. This holds especially true for rotation where very low rotational tolerances contribute to very high movement times, following an almost exponential relationship. \\
\multirow{1}{*}{\textbf{RC2}} & Avoid if possible in having pointing tasks that require movements either towards or away from the view direction of the user. Our results marginally showed that front-to-backwards movements take significantly longer than left-to-right ones and vice versa. \\
\bottomrule
\end{tabular}
\end{scriptsize}
    \vspace{-1.0mm}
  \caption{Summary of main results and key implications on object interaction in VEs and simulated teleoperation tasks.}
  \label{table:LiteratureSummary}
    \vspace{-6.5mm}
\end{table*}

\subsection{\textbf{Interpreting Results on Combined Movements in 3D}}
For combined movements, even by adding the \(IDs\) of translation and rotation (Stoelen and Akin \cite{doi:10.1177/0018720810366560}, and Kulik et al. \cite{8998368} in 2D tasks), the model is insufficient. The latter study observed only a linear fitting of $r^2=.780$ when combining both translation and rotation in simple 2D settings following movements across one line. Yet direct comparisons between different studies are difficult, particularly due to correlation coefficients ($r^2$) being highly influenced by the involved data points. Consequently, different experimental settings and manifestations do not allow for inter-study comparability \cite{10.1145/3411763.3443442, 10.1145/1753846.1753867}. Hence, we compare these models on our data directly across four experiments, as summarized in \autoref{table:ComparedMethodsFitted} and \autoref{fig:RegressionResults}.

By comparing the most widely used model extensions to date, our proposed model demonstrated better fitting results than the existing extensions both in E1 and E3, and most importantly when both translational and rotational movements were combined (E4). In experiment E4, our model was better fitting for both pointing ($r^2=.805$) and manipulation ($r^2=.803$) than any other formulation based on Fitts' law. This result can be attributed to three major factors.

Firstly, our results showed that equally weighing both \(IDs\) of rotation and translation and by extent assuming that both have a linear effect on \(MT\), is not supported by our analysis. Instead, it is crucial to separate these two terms, which can be accomplished by introducing different constants for each as explained in Section \ref{sec:modelDerivation}. Secondly, it shall be noted that the rotational tolerance $\omega$ has a non-linear relationship with \(MT\) in our study, in contrast to a linear relationship as in \cite{doi:10.1177/0018720810366560}. This difference implies that higher degrees of difficulty require matching the rotation of an object in all 3D axes, instead of one as in their work, limited to 2D space \cite{doi:10.1177/0018720810366560}. Thirdly, we can also conclude that from the 2D formulations, Hoffmann's formulation \cite{doi:10.1080/00140139508925153} indeed works well and indicates the necessity of adding the size of the object $F$, hence why it is included in our model. We can thus infer that both $W$ and $F$ inversely affect \(MT\) and should therefore be included in a model.

Our experiments found no evidence to support the significance of either directional or inclination angles in a consistent manner. More specifically we observed that directional angles presented only a significant influence towards \(MT\) limited in E2 and only in pointing ($p<.05$). Furthermore, while inclination angles affected \(MT\) both for pointing and manipulation in E2 ($p<.05$), they presented an insignificant effect in E4 with combined movements. In E4, the translational and rotational variables influenced \(MT\) significantly more than inclinations or directions. Nevertheless, despite its negligible effect on \(MT\), our study shows marginally worse performance when objects were presented along the view axis i.e. front-to-back ($90^{\circ}$, $270^{\circ}$) rather than the lateral i.e. left-to-right  ($0^{\circ}$, $180^{\circ}$). This is in line with \cite{10.1145/3290605.3300437}.

Finally, manipulation tasks took a longer time to complete in all cases except in purely rotational tasks (E3), where pointing proved significantly more difficult, especially with the lowest target tolerance at $\omega=2.5^{\circ}$. Overall, this time difference can be partially explained by the duration spent on grasping the object. An important finding was that manipulation can be modelled under Fitt's model if the important variable of object size \(F\) is taken into account. The object size is mostly overlooked in the state of the art but is mitigated in Hoffmann's work \cite{doi:10.1080/00140139508925153}, as seen in Eq. \ref{eqn:hoffman}. 

In \autoref{table:LiteratureSummary}, we summarize the main findings from our work and the implications on task design involving pointing tasks, 3D user interfaces as well as the manipulation of objects in 3D space, as seen in VR and robotic teleoperation.

\subsection{\textbf{Limitations and Future Work}}
Some limitations to overcome in a future study is to consider the multitude of different input devices and different grasping types one can use, which may significantly influence the formulation of human models \cite{10.1145/3411763.3443442}. Furthermore, human perception is a complex phenomenon and subject to each individual's exposure to technologies as well as personality-related factors, yet accounting for these is particularly challenging \cite{10.1145/3411763.3443442}.

In this work, we investigated seven distinctive spatial variables to explain and model combined translational and rotational variations in full 3D space, as shown in \autoref{table:StudyVariations}. Consequently, due to the significant amount of variations introduced, we limited the investigation of each variable to a maximum of four levels. Future studies are advised to investigate an even wider range of these variables with more levels, to further increase our understanding of their contribution towards task difficulty. For example, changing the cube shape in our study, with a more complex 3D shape such as a toy car, will allow for a wider range of rotational variations to be investigated. When the object to be manipulated is non-rigid, fragile or deformable, it will place significant influences on a 3D metric for teleoperation \cite{wen2020force}.

As for all human performance models deriving from Fitts' law, the main limitations are stationary pointing or manipulation tasks. All models studied in this paper do not take into account upper-body movements, such as movements from torsos and/or shoulders. Furthermore, biomechanical variations in participants, such as placing objects near the dominant hand or the size of the controlled end-effector are all determining factors for a model.

\section{CONCLUSION}
In this work, we investigated a new human performance metric in full 3D for both pointing and manipulation with combined translational and rotational movements. We conducted four experiments, each adding progressively higher spatial complexity. In the most basic form of 3D translational pointing and manipulation (E1), we observed that existing approaches can be used to adequately model human performance, but were insufficient when spatial arrangements were introduced (E2), such as inclinations and directions. However, the majority of these approaches did not model well the rotational movements as shown in E3 and were insufficient when combining translation and rotation with spatial arrangements as studied in E4.

To the best of our knowledge, this study extensively compared the most widely used performance metrics based on Fitts' law. We proposed a new performance model that can model human performance in full 3D space in a simulated teleoperation VR setting with two types of interactions. Our metric modelled human performance better than existing formulations, especially in increasing spatial complexities such as those seen in full 3D space.

However, numerous aspects should further be investigated and improved, such as a more comprehensive range of spatial variations (e.g. additional levels of rotations, spatial directions etc.), more complex shapes (non-primitive 3D objects), different grasping types, hand end-effectors as well as input devices. Nonetheless, this work was a first attempt towards a higher dimensional formulation to evaluate full-3D movements, which can be used in accessing human performance in a VR-based teleoperation setting and with the potential to be applied in robot shared control.

\section*{ACKNOWLEDGMENTS}
Supported by the  EPSRC Future AI and Robotics for Space (EP/R026092/1), EPSRC CDT in RAS (EP/L016834/1), and H2020 project Harmony (101017008). Ethics Approval 2019/22258. We also thank Iordanis Chatzinikolaidis for his input.

\bibliographystyle{IEEEtran}
\bibliography{references}
\end{document}